\documentclass{article}
\usepackage{graphicx} 
\usepackage{booktabs}
\usepackage{subcaption}
\usepackage{amsmath}
\usepackage{xcolor}
\usepackage[a4paper, total={6.7in, 10.5in}]{geometry}
\usepackage{fancyhdr, lipsum}
\usepackage[%
    colorlinks=true,
    pdfborder={0 0 0},
    linkcolor=blue
]{hyperref}
\usepackage{todonotes}
\usepackage{upgreek}
\usepackage{comment}
\usepackage[parfill]{parskip}
\usepackage{overpic}

\usepackage{url}
\usepackage{bm}
\usepackage{biblatex}
\newcommand{\boldu}{\ensuremath \mathbf{u}}
\newcommand{\boldv}{\ensuremath \mathbf{v}}
\newcommand{\boldx}{\ensuremath \mathbf{x}}
\newcommand{\boldX}{\ensuremath \mathbf{X}}
\newcommand{\boldn}{\ensuremath \mathbf{n}}
\newcommand{\boldvarepsilon}{\ensuremath \bm\upvarepsilon}
\newcommand{\boldpsi}{\ensuremath \bm\uppsi}
\newcommand{\boldtheta}{\ensuremath \bm\uptheta}
\newcommand{\boldsigma}{\ensuremath \bm\upsigma}

\newcommand{\boldS}{\ensuremath \mathbf{S}}
\newcommand{\boldE}{\ensuremath \mathbf{E}}
\newcommand{\bolde}{\ensuremath \mathbf{e}}

\newcommand{\ujump}{\ensuremath {[\![\boldu]\!]}}

\newcommand{\footremember}[2]{%
	\footnote{#2}
	\newcounter{#1}
	\setcounter{#1}{\value{footnote}}%
}
\newcommand{\footrecall}[1]{%
	\footnotemark[\value{#1}]%
}

\title{On representation of macroscopic crack in periodic fine-scale \\ discrete mechanical models}
\author{Jan Raisinger\footremember{Brno}{Brno University of Technology, Faculty of Civil Engineering, Brno, Czechia}\footremember{email}{Corresponding author: jan.raisinger@vut.cz} and Jan Eliáš\footrecall{Brno}}
\date{}

\begin{document}

\maketitle

\section*{Abstract}
The evolution of strain localization and the corresponding macroscopic response obtained in multiscale modeling of softening heterogeneous materials strongly depend on the boundary conditions imposed on the fine-scale model. For rectilinear models (e.g., squares or cubes), employing standard Periodic BCs leads to overestimation of energy dissipation and artificially ductile behavior, whenever the localization band orientation deviates from the periodic directions. Recently proposed Tessellation and Percolation-path-aligned BCs promise to address this by aligning the periodicity frame with evolving localization bands. Alternatively, spherical/circular models provide an~orientation independent response by design. Unfortunately, the standard Periodic BCs do not allow development of proper localization band crossing the spherical model's boundaries. A~recently proposed modification addresses this by adding a~displacement jump to the spherical periodic BCs.

This study evaluates the applicability of these novel BCs to a~mesoscale discrete particle model of concrete. Two-dimensional square and circular models under uniaxial tension with different loading directions are analyzed, with selected approaches extended to three-dimensional cube models. 
Percolation-path-aligned BCs were found to suffer from several drawbacks. Uneven straining of the weakly and strongly constrained sections can trigger localization into multiple bands, and the weakly constrained section is susceptible to spurious strain localization. Tessellation BCs, on the other hand, consistently allow the development of a~single localization band of predictable length, determined only by the model geometry, enabling a~straightforward correction of its influence during post-processing. As a~result, variations in dissipated energy with loading direction can be attributed solely to model geometry rather than to artifacts introduced by the boundary conditions.
Periodic boundary conditions augmented with a~displacement jump applied to a~circular model yield inconsistent results. Some cases exhibit correct localization band development, but some produce crack patterns similar to those under the standard Periodic BCs. Which of these occurs is dictated by the internal heterogeneity of the model, which effectively determines whether strain localization is initiated in the boundary segment enhanced by the displacement jump.

\section*{Keywords}
periodicity, boundary conditions, discrete lattice model, homogenization, strain localization

\section{Introduction}
The computational analysis of large-scale structures made of complex heterogeneous materials can be significantly facilitated by multiscale modeling techniques. Instead of explicitly resolving the entire microstructure or mesostructure, these methods utilize a~representative volume element (RVE), a small fine-scale domain whose response adequately characterizes the behavior of the bulk material. While the existence of an RVE is well established for elastic and hardening materials, its applicability to softening materials remains controversial. The main reason is the strain localization, which violates the assumptions of classical homogenization and gives rise to a deterministic size effect \parencite{GITMAN20072518}.

Despite these theoretical concerns, multiscale approaches have been successfully applied to materials exhibiting strain softening. This is demonstrated by numerous studies employing both continuous \emph{fine-scale} models \parencite{BelSon10,BosKou-14,BocKou-15,TurHoo-18,SinMah19,KeMee-22} and discrete models \parencite{EliCus26,EliCus22,RezCus16,RezZho-17,ZhuBriFas25}. These approaches use the homogenized response of a~periodic fine-scale model as the~constitutive model for the~coarse scale, treating the~fine-scale model as though it were a~classical RVE. An unresolved aspect of multiscale analysis of softening materials concerns the~choice of the~fine-scale model size, as conventional RVE-based principles cannot be employed. In Ref.~\parencite{GITMAN2008302} this is addressed through a~coupled-volume approach in which the~size of the~fine-scale model is prescribed by the~size of the associated coarse-scale finite element.
Another approach by \textcite{Ung13} decomposes the~displacement field of the~fine-scale model into a~homogeneous part, fluctuations, and a~cracking part based on additional degrees of freedom accounting for crack opening and sliding and extends the Hill-Mandel lemma so that the~fine-scale model is not required to have the~same size as the~corresponding macroscopic integration point. References \parencite{TorSan-14,OliCai-15} use similar concepts.

Independent of the issues concerning fine-scale model size and representativeness, the influence of the imposed boundary conditions (BCs) presents an important challenge. They influence the onset, evolution, and final pattern of the strain localization, affecting the homogenized response. 
In specific fine-scale models with a low number of possible localization patterns predetermined by weak zones in which fracture appears and propagates, traditional boundary conditions can yield satisfactory results. Examples include a masonry fine-scale model with periodic boundary conditions \cite{masonry} and a concrete fine-scale model with cohesive bands at predefined position constrained by minimal kinematic boundary conditions \cite{OliCai-15}.
However, for fine-scale models with complex internal and boundary structures that allow diverse localization patterns to develop,
the periodic boundary conditions, commonly adopted in homogenization simulations, lead to artificially increased  ductility of the homogenized response whenever the band's  normal direction is not parallel to the periodicity direction. This has motivated developments of alternative definitions of boundary conditions which have been investigated in several studies employing continuous fine-scale models \parencite{CoeKouGee12, goldmann2018boundary, OliCai-15}.

In this work, the following three novel approaches are investigated.
(i) The Percolation-path-aligned BCs introduced by \textcite{CoeKouGee12} and (ii) the Tessellation BCs developed by \textcite{goldmann2018boundary} are suitable for cubiodal fine-scale models. The former is based on rotation of the periodicity frame, while the latter is based on its shifting. Due to the cuboidal shape, its fracture response is inherently orientation dependent, even if the boundary conditions allow arbitrary strain localization. As a~remedy, a~circular/spherical fine-scale model can be used instead~\parencite{circularRVEorig}.  However, traditional periodic boundary conditions forbid fully developed strain localization in circular models, as there is no unique periodicity direction. (iii) To solve this, a~modification to the Periodic BCs was proposed by \textcite{sperical1}, introducing an~unknown displacement jump between two halves of the circle, based on the approach described in Ref.~\parencite{OliCai-15}.

An essential input for all of these methods is the orientation of the developing localization band.
Since generally this orientation is not known a~priori, it must usually be determined as part of the nonlinear solution procedure. Here, the problem is simplified by considering fine-scale models loaded by uniaxial tension with fixed direction, and the localization band is assumed to develop in a normal direction to the load.

The novel approaches, which were originally proposed for continuous two-dimensional fine-scale models, are herein implemented for discrete mechanical models. A 2D numerical study is performed to compare results obtained using the~novel BCs with those calculated with traditional BCs, namely the~Periodic and Minimal kinematic BCs. The differences between square and circular model geometries are also discussed. Fundamental issues of the Percolation-aligned and the periodic BCs modified with displacement jump are revealed. Their causes are discussed, with a modified variant of the former approach introduced to demonstrate the mechanism.
Following the results of the 2D study, the Percolation-path-aligned and the Tessellation BCs approaches are extended to 3D. This extension exposes several implementation challenges, which are described in the Appendices. Finally, a 3D numerical study analogous to the 2D study is conducted.

\section{Homogenization of heterogeneous Cauchy continuum \label{sec:homogenization}}
The asymptotic expansion homogenization, pioneered by \textcite{Ben78} and \textcite{San80}, introduces two different length scales. $D$ is the structural size, often taken as the maximum distance of two points within the domain; $d$ is the size of the period fine-scale model. The homogenization explores the asymptotic limit for which the ratio between $d$ and $D$ (also known as separation of scales constant) approaches zero: $\eta = d/D \rightarrow 0$. There are also two different reference systems: the coarse-scale one denoted $X$ and the fine-scale one $x$. The spatial derivative changes to $\nabla\rightarrow \nabla_X + \eta^{-1}\nabla_x$ and all variables $\bullet(\boldx)$ become dependent on both macro and micro reference systems $\bullet(\boldX,\boldx)$.

All variables are, in the limit $\eta\rightarrow 0$, expanded into infinite series with terms with increasing power of $\eta$. For example, the displacement field becomes
\begin{align}
\boldu = \boldu^{(0)} + \eta \boldu^{(1)} + \dots
\end{align}
The strain tensor is the symmetric part of the displacement gradient $\boldvarepsilon=\nabla\stackrel{\mathrm{sym}}{\otimes}\boldu$. Combining the two-scale differentiation $\boldvarepsilon=\nabla_X\stackrel{\mathrm{sym}}{\otimes}\boldu+\eta^{-1}\nabla_x\stackrel{\mathrm{sym}}{\otimes}\boldu$ with the expanded displacements yields
\begin{align}
\boldvarepsilon = \eta^{-1}\underbrace{ \left(\nabla_x\stackrel{\mathrm{sym}}{\otimes}\boldu^{(0)}\right)}_{\boldvarepsilon^{(-1)}} + \eta^{0}\underbrace{\left(\nabla_X\stackrel{\mathrm{sym}}{\otimes}\boldu^{(0)} + \nabla_x\stackrel{\mathrm{sym}}{\otimes}\boldu^{(1)}\right)}_{\boldvarepsilon^{(0)}}  + \dots
\end{align}

The expansion of the stress field follows $\boldsigma = \eta^{-1}\boldsigma^{(-1)} + \eta^{0}\boldsigma^{(0)} + \dots$. Defining the constitutive function $f_s$ and the state variables $\boldpsi$, the first term reads
\begin{align}
\eta^{-1}\boldsigma^{(-1)} = f_s\left(\eta^{-1}\boldvarepsilon^{(-1)}, \boldpsi\right)
\end{align}

The balance equation in the two scale expanded framework reads 
\begin{align}
\nabla_X \cdot \left(\eta^{-1}\boldsigma^{(-1)} + \eta^{0}\boldsigma^{(0)} + \dots\right) + \eta^{-1}\nabla_x \cdot \left(\eta^{-1}\boldsigma^{(-1)} + \eta^{0}\boldsigma^{(0)} + \dots\right) + f = 0
\end{align}
The terms containing the lowest power of $\eta$ provide the following balance equation $\eta^{-2}\nabla_x \cdot \boldsigma^{(-1)} = 0$. Its solution under the $x$-periodicity of the displacement field gives $\boldsigma^{(-1)}=\bm{0}$, $\boldvarepsilon^{(-1)}=\bm{0}$, and constantness of $\boldu^{(0)}$  in $\boldx$ reference system. We can therefore write the next stress term and proceed to the next balance equation. 
\begin{align}
\boldvarepsilon^{(0)} &= \nabla_X\stackrel{\mathrm{sym}}{\otimes}\boldu^{(0)} + \nabla_x\stackrel{\mathrm{sym}}{\otimes}\boldu^{(1)} & \boldsigma^{(0)} &= f_s\left(\boldvarepsilon^{(0)}, \boldpsi\right) & \eta^{-1}\nabla_x \cdot \boldsigma^{(0)} &= 0
\end{align}
These equations describe the fine-scale problem to be solved numerically after adding a~constraint defined by the periodic boundary conditions. This fine-scale model is the~standard mechanical model loaded  by the negative coarse-scale strain tensor $\hat\boldvarepsilon = -\nabla_X\stackrel{\mathrm{sym}}{\otimes}\boldu^{(0)}$ as eigenstrain. The solution to this problem are fluctuating displacements $\boldu^{(1)}$ emerging from the heterogeneity. They must be zero on average. The periodicity of the fine-scale model solution can be enforced either``strongly'', if the boundary nodes have periodic pairs, or by using a~weak periodicity scheme in the case of non-periodic boundary meshes \parencite{weak_periodicity}.

The next balance equation gives a~macroscopic formulation corresponding to Cauchy continuum. It also provides the macroscopic stress tensor definition
\begin{align}
\boldS &= \frac{1}{V}\int_{\Omega} \boldsigma^{(0)} \mathrm{d}\boldx \label{eq:stress_continuum}
\end{align}
where $\Omega$ is the domain of the fine-scale problem, $V$ is its volume, and $\Gamma$ is the boundary. The unit outward normal will be denoted $\boldn$. 

This article is focused to the fine-scale model only, therefore the notation is simplified by dropping the indices. The fine-scale fluctuation displacement $\boldu^{(1)}$ will be written as $\boldu$, the fine-scale strain $\boldvarepsilon^{(0)}$ will be represented by plain $\boldvarepsilon$, and the fine-scale stress $\boldsigma^{(0)}$ will become $\boldsigma$. The coarse-scale strain $\nabla_X\stackrel{\mathrm{sym}}{\otimes}\boldu^{(0)}$ will  be denoted $\boldE$.

\subsection{Periodic boundary conditions}
The fine-scale model's boundary $\Gamma$ is partitioned into two segments: $\Gamma^+$ and $\Gamma^-$. The periodic boundary conditions (PBCs) can be applied if each point $\boldx_k^-\in\Gamma^-$ is mapped to a~point $\boldx_k^+\in\Gamma^+$ by a~bijective mapping $\chi$ \parencite{goldmann2018boundary}
\begin{align}
\boldx_k^+=\chi(\boldx_k^-):\ \boldn_k^+ = -\boldn_k^-,\,\mathrm{d}A_k^+ = \mathrm{d}A_k^- 
\label{eq:periodic_condition}
\end{align}
Here, the $\mathrm{d}A$ symbols refer to the boundary segments associated with the corresponding boundary nodes. For every corresponding pair of boundary nodes, the fine-scale displacement vectors must coincide.
\begin{align}
\boldu(\boldx_k^+)=\boldu(\boldx_k^-)
\label{eq:periodic_pairs}
\end{align}
In addition, rigid-body translation of the entire fine-scale domain must be eliminated, for instance by enforcing zero average displacement or by fixing the translations of a single arbitrarily selected node.

Instead of imposing the macroscopic strain $\boldE$ by eigenstrain, one can change equation~\eqref{eq:periodic_pairs} so that the periodic boundary node pairs must move relative to each other by vector  $\boldE \cdot(\boldx^+ - \boldx^-)$. The updated equation~\eqref{eq:periodic_pairs} then reads
\begin{align}
\boldu(\boldx_k^+)=\boldu(\boldx_k^-) + \boldE \cdot(\boldx_k^+ - \boldx_k^-)
\label{eq:periodic_pairs2}
\end{align} 
This approach is equivalent to periodic BCs with eigenstrains, and is sometimes also called periodic boundary conditions. All the equations hereinafter that refer to some form of periodic boundary conditions might be actually implemented as Eq.~\eqref{eq:periodic_pairs} with eigenstrains or Eq.~\eqref{eq:periodic_pairs2} without them.

\subsection{Minimal kinematic boundary conditions}
The Minimal kinematic boundary conditions (hereafter referred to as Minimal BCs where appropriate) are derived from the requirement that the average strain over the fine-scale domain should be zero \parencite{minimalBC}
\begin{align}
\frac{1}{V}\int_\Omega \boldvarepsilon(\boldx) \mathrm{d}\boldx = \bm{0}
\label{eq:minim_strain_V}
\end{align}
Defining the fine-scale strain as the symmetric part of the displacement gradient $\boldvarepsilon=\nabla_x\stackrel{\mathrm{sym}}{\otimes}\boldu$, Equation~\eqref{eq:minim_strain_V} can be transformed, through application of the divergence theorem, into an integral over the fine-scale domain boundary $\Gamma$
\begin{align}
\frac{1}{2V}\int_\Gamma \boldu\otimes\boldn + \boldn\otimes\boldu \,\mathrm{d}\Gamma = \bm{0}
\label{eq:minim_strain_S}
\end{align}

When applied to a~rectangular or cuboid domain, each pair of opposite sides $A$ and $B$ must obey the following conditions
\begin{align}
\int_{A} u_i \mathrm{d}\Gamma - \int_{B} u_i \mathrm{d}\Gamma  &= 0
\label{eq:mkbc}
\end{align}
for $i\in\left\{1,\,\dots,\,n_d\right\}$ ($n_d$ is the number of dimensions). In addition, using the same approach as in the periodic boundary conditions, the rigid body translation of the whole fine-scale domain needs to be prevented.

\subsection{Percolation-path-aligned boundary conditions}
\begin{figure}
\centering
\includegraphics[width=6.3in]{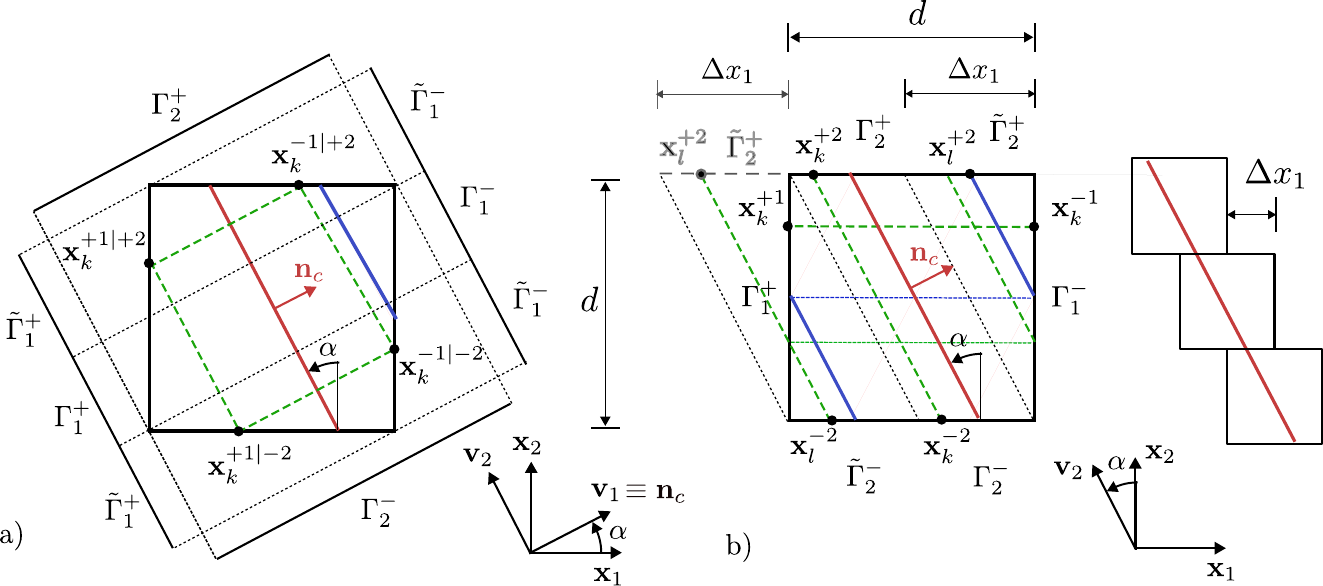}

\caption{a) Projection of the boundaries of the fine-scale model in the through-crack direction $\boldv_1$ and parallel-to-crack direction $\boldv_2$ in the Percolation-path-aligned BC formulation; b) Mapping offset $\Delta x_1$ and the associated parallel-to-crack projection in the Tessellation BCs for crack angle $\alpha<\pi/4$; red and blue lines indicate crack with the same orientation angle $\alpha$ but different position within the fine-scale model}
\label{fig:percolation_tessel_scheme}
\end{figure}

The Percolation-path-aligned BCs were developed in Ref.~\parencite{CoeKouGee12}. They will be denoted as Aligned BCs hereinafter. In this formulation, the periodicity relations of the boundary sections $\Gamma^+$ and $\Gamma^-$ are applied in a coordinate system aligned with the localization pattern. Rather than using the original $\boldx_1$ and $\boldx_2$ directions, the periodic constraints are imposed along the rotated axes $\boldv_1$ and $\boldv_2$. Assuming that, for each node $\boldx_k^{1+}$, a corresponding node $\boldx_k^{1-}$ exists on the opposite boundary in the $\boldv_1$ direction, and that each node $\boldx_k^{2+}$ has a counterpart $\boldx_k^{2-}$ in the $\boldv_2$ direction, the periodic relations illustrated in Fig.~\ref{fig:percolation_tessel_scheme}a can be established. The coordinate system is rotated by an angle $\alpha$ corresponding to the localization-band orientation.
The projection of the boundary along the~$\boldv_1$ direction, subsequently denoted as the crack direction, is further divided into two types of segments: $\Gamma_1^+$ and $\Gamma_1^-$, for which the outward normal vectors of the + and - sections are oriented oppositely, and  $\tilde{\Gamma}_1^+$ and $\tilde{\Gamma}_1^-$, where they are not.
\begin{align}
\Gamma_1: \boldn_k^{1+} &= -\boldn_k^{1-} & \tilde{\Gamma}_1: \boldn_k^{1+} &\neq -\boldn_k^{1-}
\end{align}
In the through-crack direction $\boldv_1$, the periodic constraints \eqref{eq:aligned_BCs1} are imposed only on the boundary segments $\Gamma_1$. For the segments $\tilde{\Gamma}_1$, only the weak condition defined by Eq.~\eqref{eq:aligned_BCs2} is imposed. According to \parencite{CoeKouGee12}, this avoids overconstraining the fine-scale model. Along the direction $\boldv_2$, parallel to the localization band, periodicity is enforced on all corresponding boundary segments regardless of their normal vectors. 
\begin{subequations}
\begin{align}
\boldu (\boldx_k^{1+}) &= \boldu (\boldx_k^{1-}) & \mathrm{for}\  \boldx_k^{1+} \in \Gamma_1^+
\label{eq:aligned_BCs1}  \\
\int_{\tilde{\Gamma}_1^+}\boldu (\boldx_k^{1+})\mathrm{d}\Gamma &= \int_{\tilde{\Gamma}_1^-}\boldu (\boldx_k^{1-})\mathrm{d}\Gamma & \mathrm{for}\  \boldx_k^{1+} \in \tilde\Gamma_1^+
\label{eq:aligned_BCs2}  \\
\boldu (\boldx_k^{2+}) &= \boldu (\boldx_k^{2-}) & \mathrm{for}\ \boldx_k^{2+} \in \Gamma_2^+
\label{eq:aligned_BCs3} 
\end{align}
\end{subequations}
In practice, the assumption that every node possesses a corresponding periodic image is generally not satisfied. As a result, additional procedures, described in Appendix~\hyperref[appendixA]{A}, are required to formulate the constraints for arbitrary nodal arrangements. Furthermore, rigid-body motion of the fine-scale model must be eliminated, which is achieved by constraining the translational degrees of freedom of a randomly selected node.

The extension to three dimensions is straightforward. However, while both directions $\boldv_1$ and $\boldv_2$ are uniquely defined by the crack normal $\boldn_c$ in two dimensions, in 3D, on the other-hand, a~crack with normal $\boldn_c$ only defines the $\boldv_1$ direction; the~other orthonormal directions must be chosen from an~infinite number of options. The procedure for identifying optimal parallel-to-crack directions used in this work is detailed in the Appendix~\hyperref[appendixB]{B}.

\subsection{Tessellation boundary conditions}
The Tessellation boundary conditions were introduced by \textcite{goldmann2018boundary}. Unlike Percolation-path-aligned BCs, the Tessellation BCs ensure gapless tiling of a~space by the deformed fine-scale domains. This is achieved by shifting adjacent domains by an offset $\Delta x$, enabling the localization band to continue seamlessly across domain boundaries while avoiding the artificial crack-pattern repetition imposed by standard periodic BCs. This is depicted in Fig.~\ref{fig:percolation_tessel_scheme}b. For a square two-dimensional domain, the offset $\Delta x_1$ or $\Delta x_2$, is defined by the localization-band inclination angle $\alpha$ and model size $d$.
\begin{align}
\Delta x_1&=\begin{cases} d/\tan \alpha & \alpha\in(-\pi/4, \pi/4) \\ 0 & \mathrm{otherwise} \end{cases} &
\Delta x_2&=\begin{cases} 0 & \alpha\in(-\pi/4, \pi/4) \\ -d/\tan \alpha & \mathrm{otherwise} \end{cases} \label{eq:tessel_delta}
\end{align}
As in the case of the Percolation-path-aligned BCs, the periodic relations are formulated in a crack-aligned coordinate system. For $\Delta x_1>0$, the boundary  segment $\Gamma_2$ is partitioned into two parts and the periodic pairs are defined as shown in Fig.~\ref{fig:percolation_tessel_scheme}b: $\Gamma_2$ is shifted by $\Delta x_1$ and $\tilde{\Gamma}_2$  by $\Delta x_1-d$. Then periodic boundary conditions are prescribed between the positive ($+$) and negative ($-$) sides.
\begin{align}
\boldu (\boldx_k^{1+}) &= \boldu (\boldx_k^{1-}) & \boldu (\boldx_k^{2+}) &= \boldu (\boldx_k^{2-})
\end{align}
As in the case of the Percolation-path-aligned BCs, the shifted boundary typically does not have node pairs in the inclined periodicity direction. The displacement periodicity on non-periodic nodes is again defined by the~techniques described in Appendix~\hyperref[appendixA]{A}. As before, rigid-body translation of the entire fine-scale model is eliminated by fixing the translations of a randomly selected node.

When extended to 3D, the shifting is performed in two orthogonal directions, alongside two of the three opposing cube face pairs. The pair, in which a~face's normal has the lowest direction difference to the crack normal $\mathbf{n}_c$ (evaluated via the vector product), is prescribed non-shifted periodic boundary conditions. In the other two directions, the offsets are calculated analogously to Eq.~\eqref{eq:tessel_delta} and the shifted periodic constraints are constructed using the procedure described in Appendix~\hyperref[appendixA]{A}.

\subsection{Circular fine-scale model}

\begin{figure}[tb!]
    \centering
    \includegraphics[width=3.7in]{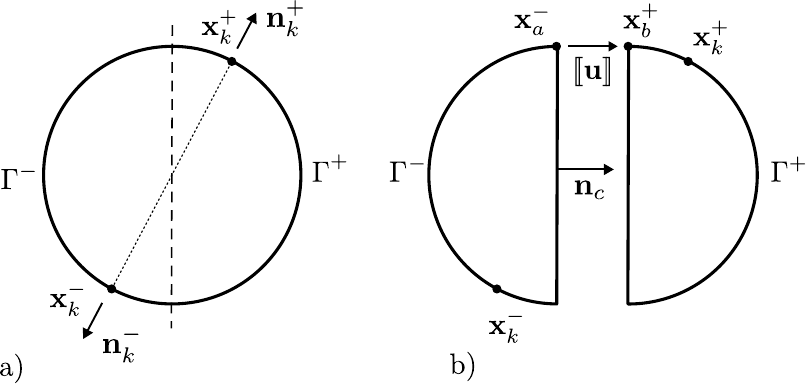}
    \caption{a) Opposing (periodic) point pair on circular fine-scale model boundary; b) the same fine-scale model with the unknown jump $[\![\boldu]\!]$ calculated as difference in displacement of the \textit{opening node pair} ($a,b$), defined by the crack normal $\boldn_{c}$.}
    \label{fig:circular_scheme}
\end{figure}

The circular fine-scale model, discussed, for example, in Ref.~\parencite{circularRVEorig}, is intrinsically orientation independent because of its shape. To apply the periodic boundary conditions, which ensure consistency between the scales, the fine-scale displacements of pairs of points with anti-periodic normals are coupled analogously to the square model (Eq.~\ref{eq:periodic_pairs} with eigenstrain or Eq.~\ref{eq:periodic_pairs2}). 

Because these boundary conditions prevent the localization band from crossing the boundary, a~modification was developed that introduces a~discontinuity in the periodicity constraints in~\parencite{sperical1}. In this scheme, upon strain localization onset, an~unknown discontinuity $\ujump$ is added to the constraint \eqref{eq:periodic_pairs} or \eqref{eq:periodic_pairs2}, defined as the relative displacement of the two nodes of the \emph{opening node pair}. The opening node pair is determined by the crack normal $\boldn_c$ and is defined as the pair of nodes located on opposite sides of the intersection between the crack surface and the boundary, as shown in Fig.~\ref{fig:2D_load_scheme_GEO}c). In 2D, either of the two intersections may be chosen arbitrarily. The addition of the jump allows for the separation of the two halves of the circular model, as illustrated in Fig.~\ref{fig:circular_scheme}b),  provided that the localization band crosses the boundary at the opening node pair. To allow unloading of the bulk material with positive $\ujump$, a~correction strain term $\hat{\boldE}$ is defined as
\begin{align}
\hat{\boldE} = \frac{2r}{|\Omega|} \ujump \stackrel{\mathrm{sym}}{\otimes} \boldn_c
\label{eq:circle_correction_strain}
\end{align}
where $|\Omega|=\pi r^2$ is the area of the circular model with radius $r$, $2r$ is the effective crack length. These augmented periodic boundary conditions, hereafter denoted Embedded-crack boundary conditions, take the form of the~following boundary pair constraint
\begin{align}
\boldu^+&=\boldu^- + \ujump & \boldu^+&=\boldu^- + (\boldE-\hat{\boldE}) \cdot (\boldx^+ - \boldx^-) + \ujump
\label{eq:circle_ModPB}
\end{align}
where the left-hand-side equation is updated Eq.~\eqref{eq:periodic_pairs} for which the imposed eigenstrain changes to $\boldE-\hat{\boldE}$. The~right-hand-side equation represents the second approach previously given by Eq.~\eqref{eq:periodic_pairs2}.

In the original formulation, the displacement jump $\ujump$ is introduced into the boundary conditions only at the~onset of localization, thereby requiring its explicit identification. In the present work, for simplicity, the~jump is included from the beginning of the simulations, with negligible influence on the elastic response or the onset of softening. Furthermore, in contrast to Ref.~\parencite{sperical1}, where unbreakable inclusions restricted certain crack orientations, the present approach allows the localization band to form between any boundary nodes. Consequently, the~distinction between effective and average crack normals, as introduced in Ref.~\parencite{sperical1}, is not required here.

\section{Modeling framework}\label{sec:Modeling_framework}

\begin{figure}[tb!]
    \centering
    \includegraphics[width=3.5in]{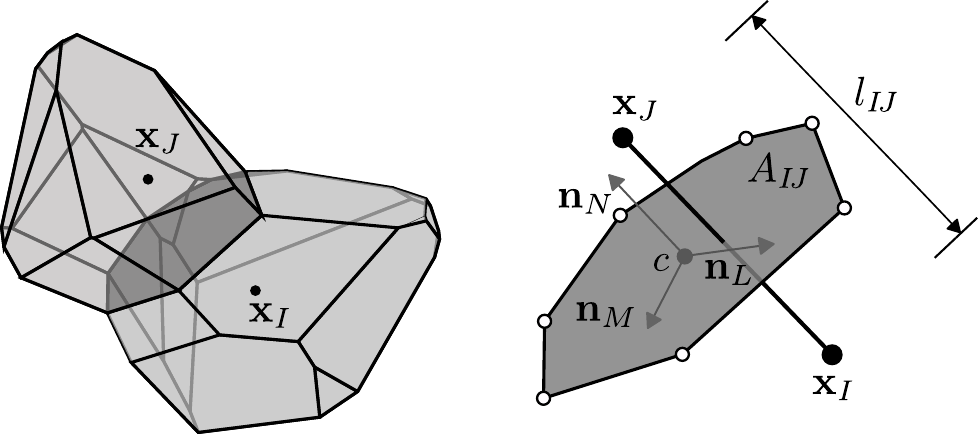}
    \caption{Two neighboring rigid particles and planar contact face with local orthonormal reference system.}
    \label{fig:face}
\end{figure}

The discrete fine-scale models, representing concrete at the mesoscale, are mechanical systems comprising interconnected rigid particles with constitutive behavior defined on inter-particle contacts~\parencite{BolEli-21}. The process of generating a~geometry of the model inside the prescribed shape (here a~square, circle, or cube) starts with random placement of circular/spherical inclusions with diameters distributed according to the Fuller curve with exponent 0.5. The placement starts with the largest inclusions, acceptance of each new one is conditioned by no overlap with previously approved inclusions. The weighted Delaunay triangulation and power tessellation on these inclusions then provides the convex polyhedral particles (rigid bodies) with movements governed by translational, $\boldu$, and rotational, $\boldtheta$, degrees of freedom defined in the centers of the inclusions. The contact between the rigid particles is characterized by the area, length, and a~local coordinate system ($N,M,L$). One such contact is sketched in Fig.~\ref{fig:face}. Such a~way of creating the model internal structure has been pioneered by~\textcite{BOL98} using the Voronoi tessellation.

A single integration point per contact is defined at the centroid of the contact face. Assuming small displacements and rotations, the rigid body kinematic equation provides strain measures ($\bm{e}$) at the integration point evaluated as discrete displacement jump between the rigid bodies divided by contact length. The vectorial constitutive model formulated in the local reference system then provides traction at the integration point. The constitutive model is a~modified version of material behavior originally introduced in Ref.~\parencite{CUSATIS20073} and later refined in the Lattice-Discrete Particle Model (LDPM) articles~\parencite{CusPelMen11,ZhuPat-22,TroLalCus25,BhaGom-21}. The exact constitutive formulation used here can be found in Refs.~\parencite{Eli16,EliCus25,RAISINGER2025111362}. The fundamental equations are completed by balance equations for linear and angular momentum for each rigid body.

Constant material parameters are used for all simulations, with values: $E_0 = \nobreak 60$\,MPa, $\alpha = \nobreak 0.2$, $f_t = \nobreak 2.8$\,MPa, $G_t = \nobreak 50$\,Jm$^{-2}$. The material parameters are homogeneous throughout the model geometry and the heterogeneity of the stress field is a~consequence of heterogeneous internal structure only. 

\subsection{Boundary geometry} \label{seq:boundary_geo}

\begin{figure}[tb!]
\centering
\begin{overpic}[width=1.6in,clip=false]{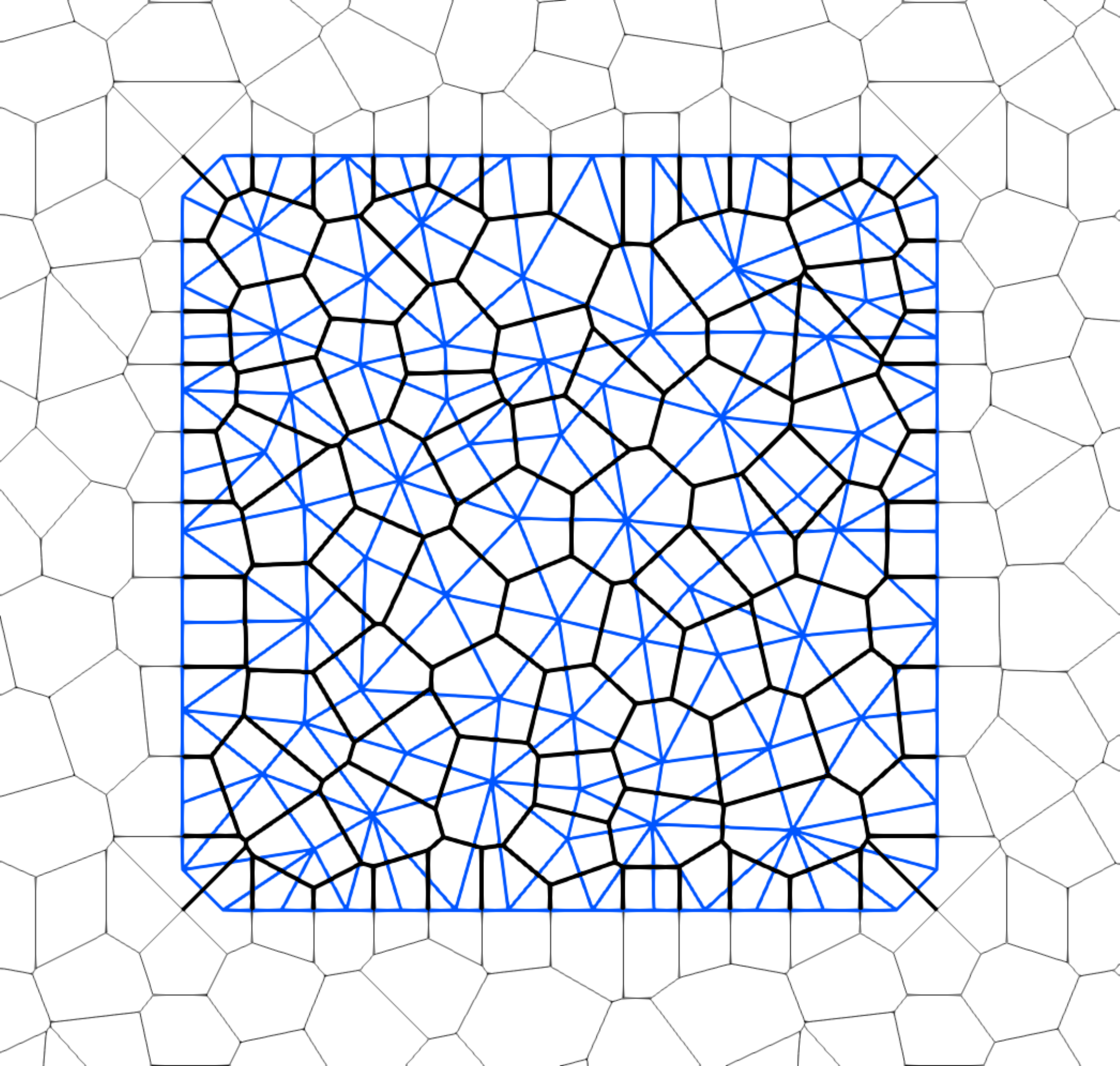}
  \put(-10,5){\text{a)}}
\end{overpic}
\hspace{0.5in}
\begin{overpic}[width=1.6in,clip=false]{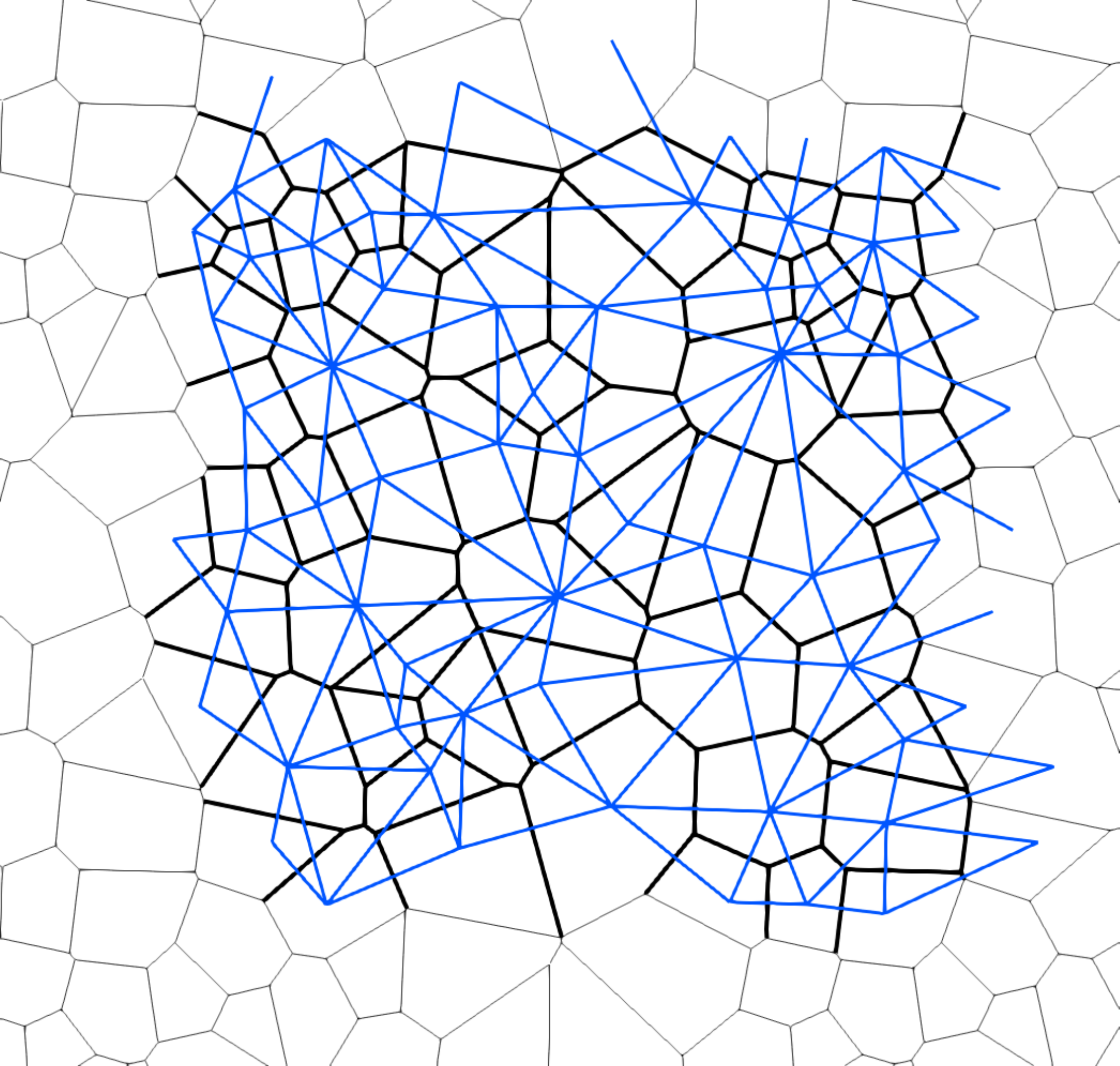}
  \put(-10,5){\text{b)}}
\end{overpic}
\hspace{0.5in}
\begin{overpic}[width=1.6in,clip=false]{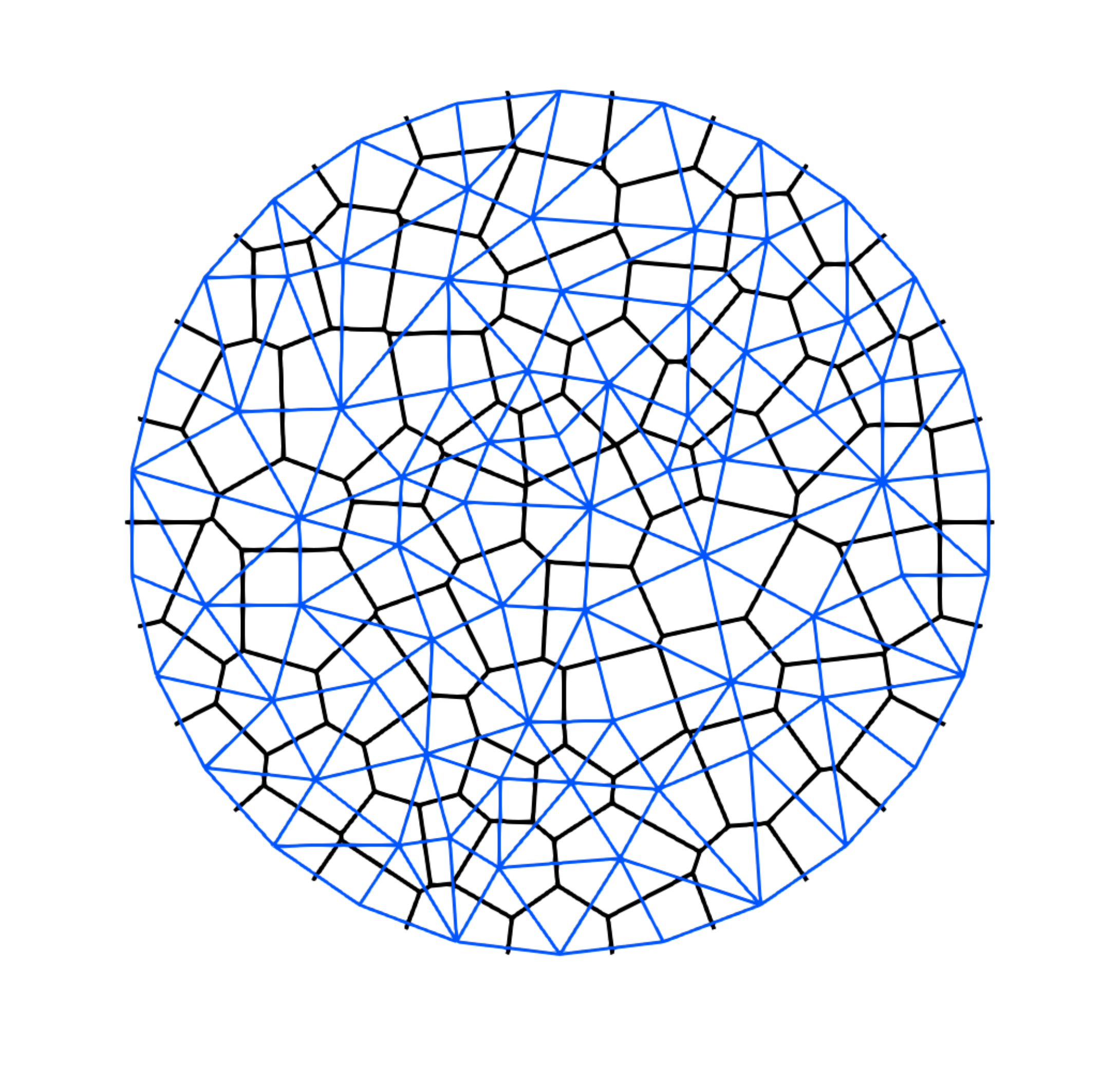}
  \put(-10,5){\text{c)}}
\end{overpic}

\caption{2D fine-scale model of a) square shape with \emph{smooth} surface (S); b) square shape with \emph{rough} surface (R); c) \emph{circular} shape with smooth surface (C); blue lines depict inter-particle elements and black lines depict contact faces.}
\label{fig:2D_GEO_sketches}
\end{figure}

The square model geometries were generated in two variants that differed in the tessellation at the boundaries. The \emph{smooth} surface (S) variant featured auxiliary nodes located irregularly at the boundaries. These nodes are placed in a~periodic manner on the opposite sides. The normal of the boundary is constant for each side of the~square domain. One such internal structure is shown in Fig.~\ref{fig:2D_GEO_sketches}a. The smooth surface results in a~boundary or wall effect~\parencite{periodic_boundary}: the boundary regions exhibit different stiffness compared to the bulk of the fine-scale model due to the structured tessellation. In contrast, models with \textit{rough} surface (R) eliminate this issue by keeping the~tessellation in the boundary region statistically identical to the bulk.
In the latter case, the placement of the particles in the~domain allows inclusions protruding from the~square domain, but restricts overlapping with the previously placed inclusions as well as with their periodic images. The power tessellation is performed on the~original inclusions and their periodic images. This results in a~periodic internal structure that is not affected by the boundaries~\parencite{periodic_boundary,EliCus22}. An~example of such \emph{rough} surface model is shown in Fig.~\ref{fig:2D_GEO_sketches}b.

The boundary nodes on the \emph{rough} surface are often poorly connected to the remaining structure and become unstable unless standard Periodic boundary conditions are applied. The strong periodicity in the rotated frame suffices as well, but the Minimal kinematic BCs and the $\tilde\Gamma_1$ boundary part of the Aligned BCs produce a~singular stiffness matrix or excessive displacements of these nodes even in elastic regimes. Additionally, the $\boldv$ projections of such boundaries lead to overlap or change of order of the boundary nodes, further preventing the use of the Aligned BCs with these models. In contrast, the Tessellation BCs can still be applied, as they enforce strong periodicity for the whole boundary. Although shifting the boundary can cause occasional overlap and change in order of the boundary nodes, particularly for large $\alpha$, this effect did not significantly influence the results for the geometries considered in the present study.

\subsection{Homogenization of discrete systems}

\textcite{RezCus16,RezZho-17} were the first to develop homogenization techniques for discrete systems. Their work was further extended in Refs.~\parencite{EliCus22,EliCus26}. The major difference from standard continuum models is the presence of rotational degrees of freedom. Nevertheless, the coarse scale still homogenizes to the Cauchy continuum under the assumption of realistic bending stiffness~\parencite{EliCus26}, while at the fine-scale the standard discrete model is used. This result follows the pioneering work of~\textcite{ForPra-01}, which showed that the Cosserat continuum at a~fine scale results in a~Cauchy continuum on a~coarse scale under the same assumption. The symmetric coarse-scale strain tensor $\boldE$ is still driving the fine-scale problem. For truly periodic boundary conditions (Eq.~\ref{eq:periodic_pairs}), it serves as an~eigen-strain tensor projected on individual discrete contacts. In the local coordinate system of each element given by the unit vectors $\boldn_{\alpha}$ ($\alpha\in\left\{N,M,L\right\}$ in 3D), this eigen-strain reads
\begin{align}
\hat e_{\alpha} &= -\boldn_N\cdot\boldE\cdot\boldn_{\alpha}
\end{align}
The expression for the coarse-scale stress tensor given in Eq.~\eqref{eq:stress_continuum} can be rewritten as
\begin{align}
\boldS &= \frac{1}{V}\sum_{e} \sum_{\alpha}l^e A^e t_{\alpha}  \boldn_{N}^e  \otimes  \boldn_{\alpha}^{e}
\end{align}
where the outer summation visits all discrete contacts within the fine-scale model.

The boundary conditions described in Sect.~\ref{sec:homogenization} are directly applicable for translational degrees of freedom. However, the BCs variants need to be extended to account for rotations as well. The periodic boundary conditions simply connect all the rotations of each dependent pair~\parencite{EliCus26}.
\begin{align}
\boldtheta (\boldx_k^+) &= \boldtheta (\boldx_k^-)
\end{align}
This strong rotational periodicity on appropriate pairs of boundary nodes is then enforced whenever strong periodicity is prescribed. In situations when only weak periodicity applies (Minimal kinematic BCs and $\tilde\Gamma_1^+$ in Percolation-path-aligned BCs) the rotations are not subjected to any constraint. Unlike translations, the average fine-scale rotation does not need to vanish~\parencite{ForPra-01,EliCus26}.

\section{Numerical results}

The numerical studies are conducted predominantly in two dimensions. Only selected cases are extended to three dimensions. All the simulations are performed in open-source software called Open Academic Solver (OAS, \url{https://gitlab.com/kelidas/OAS}). 

\subsection{Elastic properties of 2D fine-scale models}

\begin{figure}[tb!]
\centering
\includegraphics[width=6.5in]{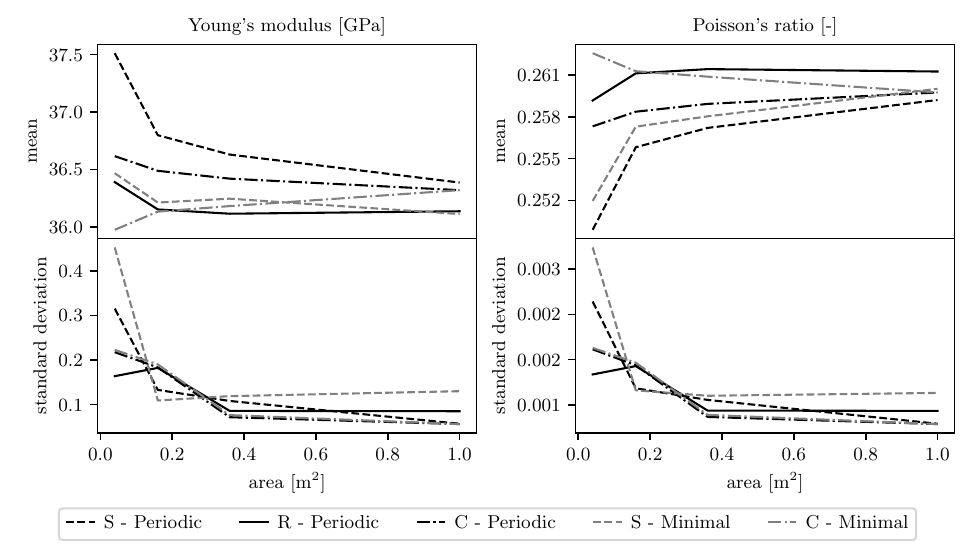}\hfill
\caption{Mean and standard deviation of effective Young's modulus and Poisson's ratio  for different sizes of 2D fine-scale models with smooth (S), rough (R) and circular (C) surface and Periodic and Minimal BCs}
\label{fig:2D_size_Enu}
\end{figure}

\begin{figure}[tb!]
\centering
\includegraphics[width=6.5in]{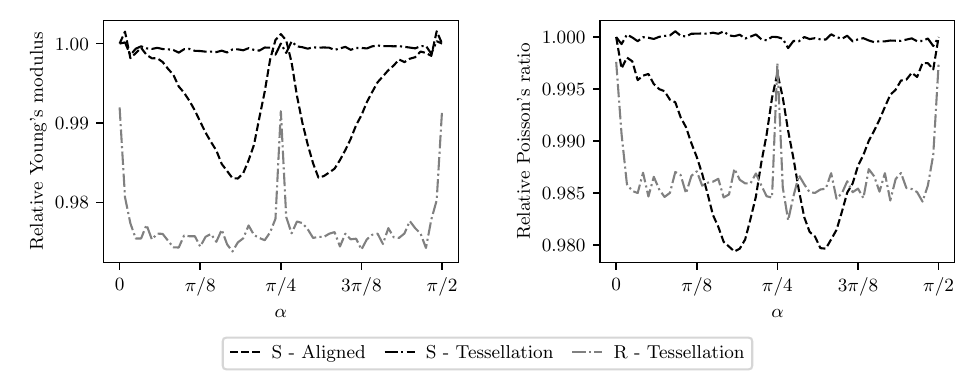}\hfill
\caption{Effective Young's modulus and Poisson's ratio normalized by their reference values dependent on the rotation of the reference frame $\alpha$ for models with smooth (S) and rough (R) surface and Aligned and Tessellation BCs}
\label{fig:2D_angle_Enu}
\end{figure}

The effective macroscopic elastic properties of the fine-scale models are known to be dependent on the boundary conditions, particularly for models with low number of grains, as documented, e.g., in Ref.~\parencite{minimalBC}. With an~increasing number of grains, achieved here by enlarging the domain while keeping the same grain size distribution, it is expected that the elastic properties converge to a~single set of values for given microscopic material parameters. 

To verify this behavior for the discrete models at hand, a~following analysis was performed: square models with \emph{smooth} (S) and \emph{rough} (R) surface were generated with side lengths of 200, 400, 600, and 1000\,mm. For each size and boundary type, 10 geometries were created. In addition, for each size, 10 \emph{circular} (C) models with the same area as the corresponding square models were generated. Then, for each model, the effective Young's modulus and Poisson's ratio were calculated as follows: the model, under purely elastic conditions, was subjected to a~set of independent macroscopic strain states (two unit axial strains and unit shear) and for each of them the homogenized stress response was calculated. These stress–strain pairs were used to assemble the homogenized matrix of elastic constants. Assuming an~isotropic macroscopic response, the effective Young’s modulus and Poisson’s ratio are determined by least-squares fitting the theoretical isotropic matrix to the numerically calculated matrix of material constants.

The effective parameters were calculated for the S and C models using the Periodic and the Minimal BCs. For the models with rough surface, the Minimal boundary conditions cannot be used, as discussed in Sec. \ref{seq:boundary_geo}. The mean values and standard deviations of coarse-scale elastic constants, computed from 10 realizations for each model size and type of boundary conditions, are shown in Fig.~\ref{fig:2D_size_Enu}. As expected, the small model sizes with Minimal kinematic BCs exhibit lower stiffness than those with Periodic BCs. This difference decreases as the model size increases. Circular and periodic boundary models show reduced variability in elastic parameters in all sizes. For small model sizes, the advantage of the R (and also C) models is evident, as they suffer the stiffness overestimation to a~much lesser degree compared to the S model. This is surprising for C models as they also have structured smooth boundary geometry.

Before conducting the fracture analysis, the novel boundary conditions in the elastic regime were verified. In the original formulation of the Embedded-crack BCs, here applied to the circular models, the displacement jump is introduced only at the strain localization onset. Therefore, in elasticity, they coincide with Periodic BCs and their influence on elastic response needs not be investigated. The Aligned and Tessellation BCs were applied to the sets of S and R models of size 200\,mm at different angles $\alpha$ from 0 to $\pi/2$. The effective elastic parameters were calculated and normalized by the \emph{reference} values calculated with the Periodic BCs. The results are shown in Fig.~\ref{fig:2D_angle_Enu}.

The Tessellation BCs on S models influence the effective elastic properties minimally compared to Periodic ones. The Aligned BCs, which can only be applied to S models, show lower stiffness for values of $\alpha$ between 0 and $\pi/4$ and $\pi/4$ and $\pi/2$. This is likely related to the lower number of constraints at these $\alpha$ values caused by projection of the borders, compared to the Periodic BCs. At $\alpha = \pi/4$, the number of constraints is very close to the Periodic BCs and the elastic constants almost coincide. The Tessellation BCs applied to the R models show lower stiffness for most of the $\alpha$ directions. This is also caused by the lower number of integration zones (and, therefore, constraints) due to the projection of the nonuniform boundary geometry. The lower values at the 0, $\pi/4$ and $\pi/2$ angles, where the Tessellation BCs should equal Periodic BCs, arise from numerical inaccuracies in the Lagrange multiplier implementation and ambiguous corner node assignment to one or the other side of the rectangular domain.

In all cases, the maximum difference is close to 2.5\,\%. Therefore, all the boundary conditions are considered to be valid for linear elasticity.

\subsection{Fracture in 2D fine-scale models}


\begin{figure}[tb!]
\centering
\includegraphics[width=6.6in]{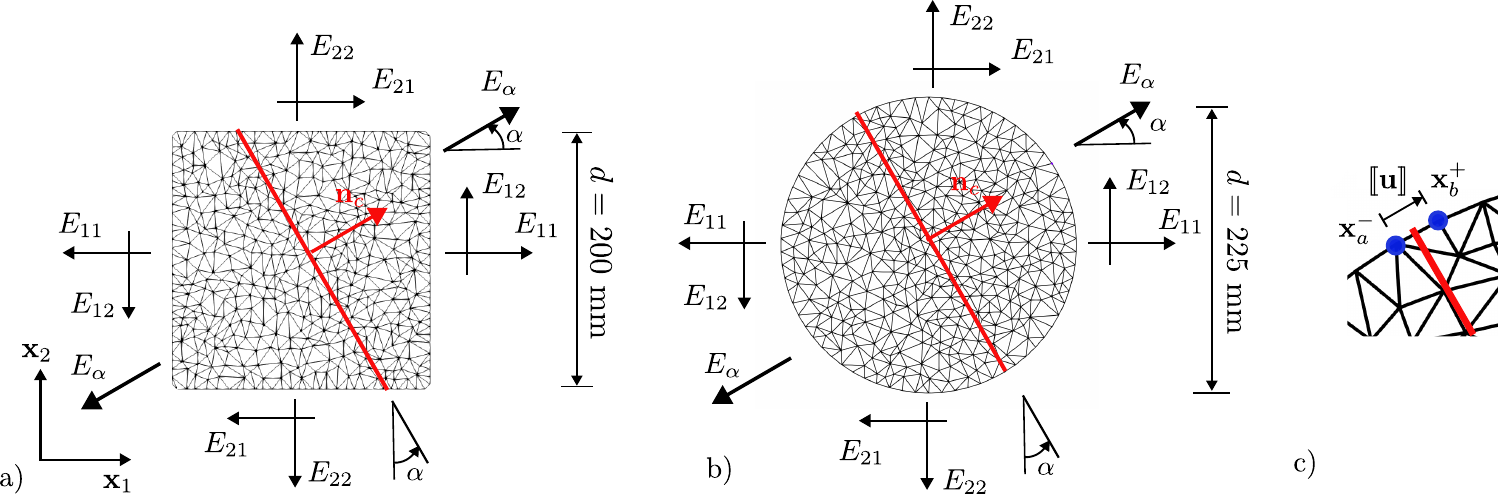}\hfill
\caption{a) Square model loaded by uniaxial tensile strain under angle $\alpha$; b) circle model loaded identically; c) detail of the opening node pair}
\label{fig:2D_load_scheme_GEO}
\end{figure}

In this section, the strain localization under different types of periodic boundary conditions is analyzed. Twenty periodic square fine-scale models with a side length of $d=200 \ \mathrm{mm}$ were generated. The individual realizations differ in the spatial distribution of the virtual circular aggregates and in the associated tessellation. In all cases, the square domain was oriented so that its sides were parallel to the $x_1$ and $x_2$ axes.

\begin{figure}
\centering
\includegraphics[width=6.5in]{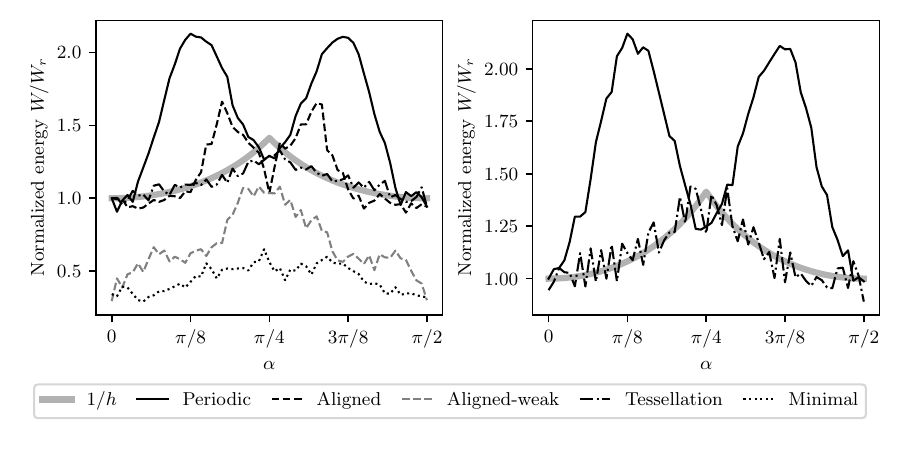}
\caption{Average relative total energy $W$ for different BC types and crack angle $\alpha$ from 20 square 2D fine-scale square models with $d = 200$\,mm: left-hand side: Smooth surface; right-hand side: Rough surface}
\label{fig:2D_energy}
\end{figure}

The models were subjected to loading through projection of the coarse-scale strain tensor $\boldE$ describing uniaxial tension in different directions $\alpha$ (Fig.~\ref{fig:2D_load_scheme_GEO}a).
The normal $\boldn_c = \boldv_1$ of the assumed crack, which defines the~Percolation-path-aligned and Tessellation BCs, was taken to be parallel to the tensile loading direction. The~principal strain $E_{I}$, the only nonzero component of $\boldE$, was increased from zero to $1.2\times 10^{-3}$ in 73 pseudo-time steps.

\begin{figure}
\centering
\includegraphics[width=6.5in]{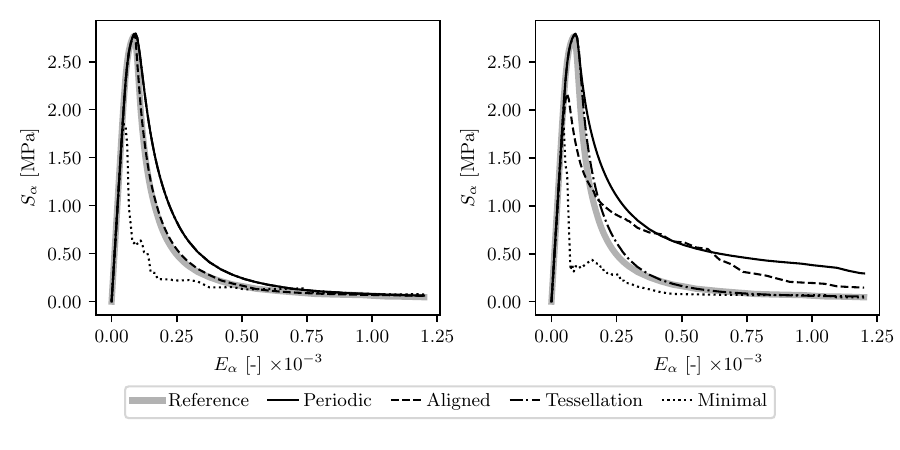}
\caption{Average stress-strain responses obtained from 20 two-dimensional square fine-scale models with side length $d=200\,\mathrm{mm}$ subjected to uniaxial tension. Results are shown for loading angles $\alpha=\pi/4$ (left) and $\alpha=0.52$ (right). The reference response corresponds to the models with Periodic BCs loaded at $\alpha=0$.}
\label{fig:2D_stress_strain}
\end{figure}

The orientation of the applied load, defined by the angle $\alpha$, was varied from $0$ to $\pi/2$ in 31 steps. These limits reflect that the problem is $\pi/2$ periodic in the statistical sense.
The homogenized coarse-scale stress $\boldS$ and the total internal energy $W$ were recorded and subsequently averaged over the 20 geometries. Average stress-strain curves for two selected loading directions are presented in Fig.~\ref{fig:2D_stress_strain}. On the horizontal axis the normal strain in the loading direction (i.e., the principal strain $E_{I}$) is plotted, while the vertical axis shows the normal stress in the same direction: $E_{\alpha} = \boldv_1\cdot \boldE \cdot \boldv_1$ and $S_{\alpha} = \boldv_1\cdot \boldS \cdot \boldv_1$ with $\boldv_1$ denoting the loading direction unit vector. The \emph{reference} curve corresponds to the average response obtained with the Periodic BCs at $\alpha=0$. For $\alpha = 0$, as well as for $\pi/4$ and $\pi/2$, the localization band can develop perpendicular to the loading direction without violating periodicity of the Periodic BCs. In the absence of boundary-condition-induced artifacts, all other loading angles and boundary-condition formulations would be expected to produce identical results.

\begin{figure}
    \centering
    \includegraphics[width=6.5in]{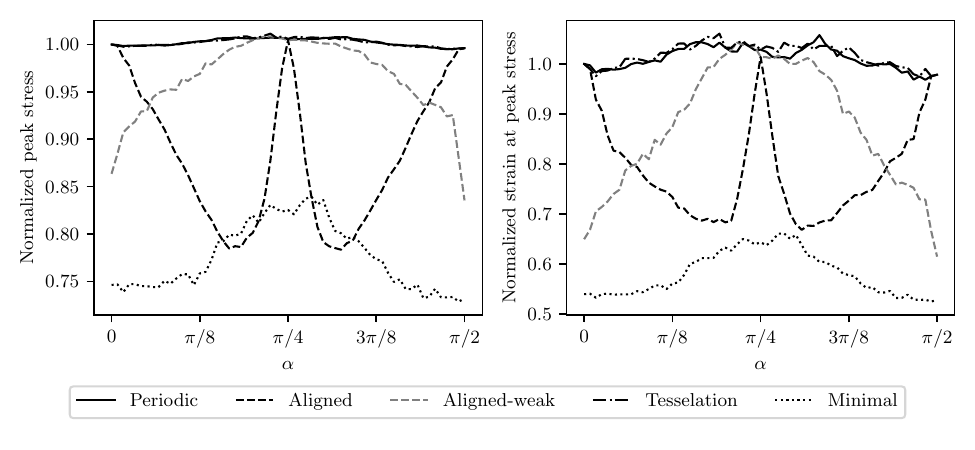}
    \caption{Average relative peak stress (left-hand side) and normalized strain at peak stress (right-hand side) for different BC types and varying crack angle $\alpha$ from 20 two-dimensional square fine-scale models of size $d = 200$\,mm}
    \label{fig:2D_stress_peak}
\end{figure}

The boundary-condition formulations are compared using three quantities: the average total energy $W$, the average peak stress $S^p_{\alpha}$, and the average strain $E^p_{\alpha}$ corresponding to the peak stress.
Because the simulations are typically terminated in state not corresponding to the stress-free crack, the reported energy includes both dissipated and residual strain-energy contributions, with the latter typically representing only a minor fraction. Relative energy values as function of the loading angle, normalized by the average energy obtained from the reference simulations, are reported in Fig.~\ref{fig:2D_energy}. Figure~\ref{fig:2D_stress_peak} shows the peak stresses and corresponding strains, also normalized by the reference model results.

\begin{figure} 
\includegraphics[width=6.5in]{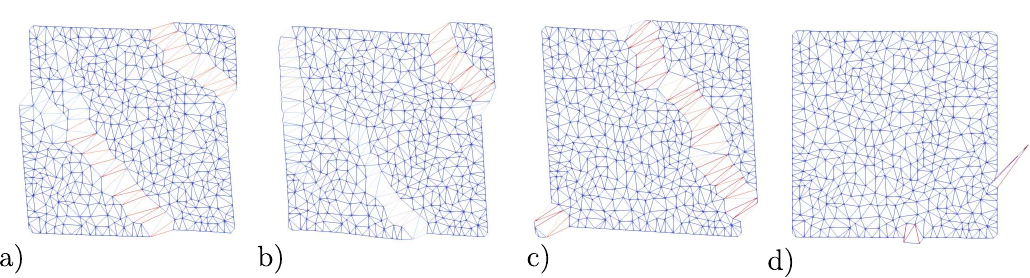}
\caption{Crack patterns in 2D square fine-scale model with  size $d=200$\,mm loaded in uniaxial tension under angle $\alpha=0.59$, with (a) Periodic, (b) Aligned, (c) Tessellation, and (d) Minimal boundary conditions. Coloring represents the crack opening, ranging from 0 (blue) to 0.19\,mm (red).}
\label{fig:2D_200_0.59_cracks}
\end{figure}

The results obtained with the \emph{Periodic BCs} are consistent with observations reported in previous studies \parencite{sperical1,jirasek}. For loading directions other than $\alpha=0,\pi/4$ and $\pi/2$, increase in the total energy is observed. This stems from the periodicity constraints, which force the formation of repeating crack patterns composed of multiple localization bands that are not orthogonal to the straining direction.
As the total crack length increases, additional energy is required for their development, resulting in a more ductile response of the fine-scale models. The effect reaches its maximum at $\alpha=\pi/8$ and $\alpha=3\pi/8$, where the total energy exceeds that of the reference case by more than a factor of two. The localization patterns exhibit spurious branching and incorrect inclination for most load orientations, as illustrated for $\alpha=0.59$ in Fig.~\ref{fig:2D_200_0.59_cracks}a. 

As noted in Refs.~\parencite{CoeKouGee12,goldmann2018boundary}, \emph{Minimal kinematic BCs} are unsuitable for problems involving strain localization. Owing to their weak kinematic constraints, localization tends to develop spuriously along the boundaries of the fine-scale domain, where detachment of a small number of boundary nodes is energetically more favorable than the formation of a localization band across the domain.
Figure~\ref{fig:2D_200_0.59_cracks}d illustrates a typical example of this behavior. The short localization band leads to low energy dissipation and low peak stress and corresponding strain compared to the reference values, as shown in Figs.~\ref{fig:2D_energy} and \ref{fig:2D_stress_peak}.

The \emph{Percolation-path-aligned BCs} were designed to accommodate a single localization band regardless of the loading orientation $\alpha$, which should result in an energy dissipation nearly independent of $\alpha$ and close to the reference value. Contrary to this expectation, significantly higher energy values were obtained for $\alpha \in (\pi/8, 3\pi/8)$. Moreover, the onset of localization exhibits a pronounced dependence on the loading orientation, reflected in the peak stresses and corresponding strains shown in Fig.~\ref{fig:2D_stress_peak}. Both quantities are markedly lower than those of the reference solution, as also demonstrated by the stress-strain responses in Fig.~\ref{fig:2D_stress_strain}. The cause is the faster crack opening in region $\Gamma_1$ with strong periodicity compared to the region $\tilde\Gamma_1$ with weak periodicity. 
The opening is driven by releasing the strain energy from unloading parts, but the size of the unloading parts differ and, hence, the energy flux. In the $\Gamma_1$ region, the size of the material domain measured in the direction $\boldn_c=\boldv_1$ is $d/\cos\alpha$, but in the $\tilde \Gamma_1$ region it is an~average value much smaller than $d$. The model either opens a~second crack in the $\Gamma_1$ region or accounts for uneven crack opening by large shear strains. Which of these situations occurs depends on the internal heterogeneity, the loading angle, and the material parameters. In the models studied, the second crack often occurred for $\alpha \in \langle\pi/8,3\pi/8\rangle$. When using a~lower value of the elastic parameter $E_0$, no secondary cracks were detected at any angle.

To substantiate the proposed explanation of the orientation dependence observed for the Aligned BCs, an alternative formulation referred to as \emph{Aligned-weak BCs} was considered. In this approach, the weak condition associated with  $\tilde{\Gamma}_1$ is applied along the entire boundary, meaning that the periodicity in the $\boldv_1$ direction is enforced only through a single integral constraint. This eliminates the occurrence of multiple localization bands and the corresponding energy increase. 
It, however, leads to spurious localization similar to that observed for the Minimal kinematic BCs when $\alpha$ is not close to $\pi/4$, as the constraints imposed on $\Gamma_2$ are insufficient to prevent it.

By construction, the \emph{Tessellation BCs} give rise to a single localization band for all values of $\alpha$. Owing to the shifted periodic constraints, this band may appear segmented as it exits and re-enters the fine-scale domain, as illustrated in Fig.~\ref{fig:2D_200_0.59_cracks}c.
The length of the localization band varies with its inclination and reaches a maximum for $\alpha=\pi/4$.
The energy dissipation is proportional to the relative length of the crack $l_c/d$. For square domains, it reads~\parencite{h_value}
\begin{align}
\frac{l_c}{d} = \frac{1}{h} = 
\begin{cases}
\dfrac{1}{\cos \alpha} & \alpha \leq \pi/4 \\[2mm]
\dfrac{1}{\sin \alpha} & \alpha > \pi/4
\end{cases}
\label{eq:h_factor2D}
\end{align}
where $h$ is the factor introduced in Ref.~\parencite{h_value} to account for this geometric effect.
Figure~\ref{fig:2D_energy} demonstrates close agreement between this factor and the normalized dissipated energy under the Tessellation BCs. Theoretically, the influence of the square domain could be corrected for during post-processing using the factor. The Tessellation BCs can be viewed as true periodic boundary conditions in a~rotated frame of reference, where the original square domain of size $d$ is reshaped to a~rectangular one with dimensions $d/h$ perpendicularly and $dh$ parallel to the crack normal.

\begin{figure}[tb!]
\centering
\includegraphics[width=3.2in]{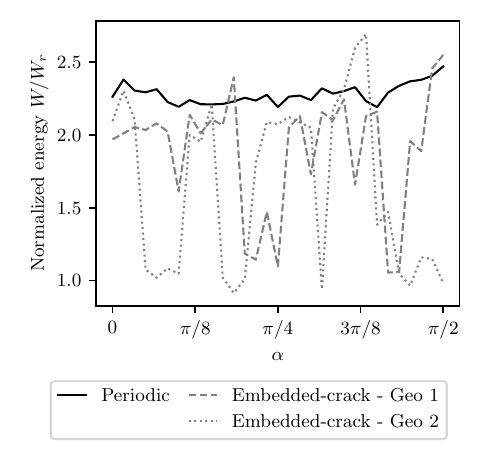}\hfill
\includegraphics[width=3.2in]{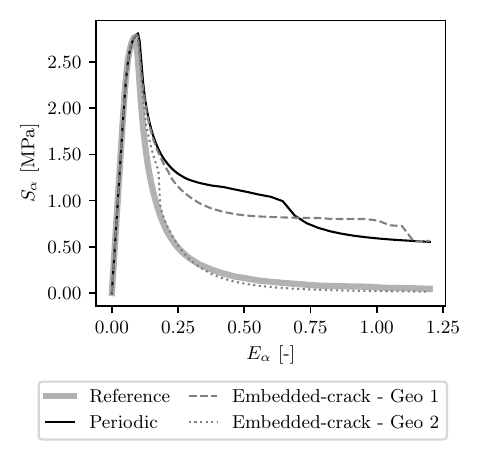}
\caption{Left-hand side: Relative energy $W$ averaged over 10 circular models with Periodic BCs compared to 2 different circular models with the Embedded-crack BCs; right-hand side: Stress-strain diagram showing the difference in response of two circular models to the same load, depending on whether the localization happens between the opening nodes (Geo 2) or not (Geo 1).}
\label{fig:2D_circle_energy_LD}
\end{figure}

\begin{figure}[tb!]
\centering
\includegraphics[width=6.5in]{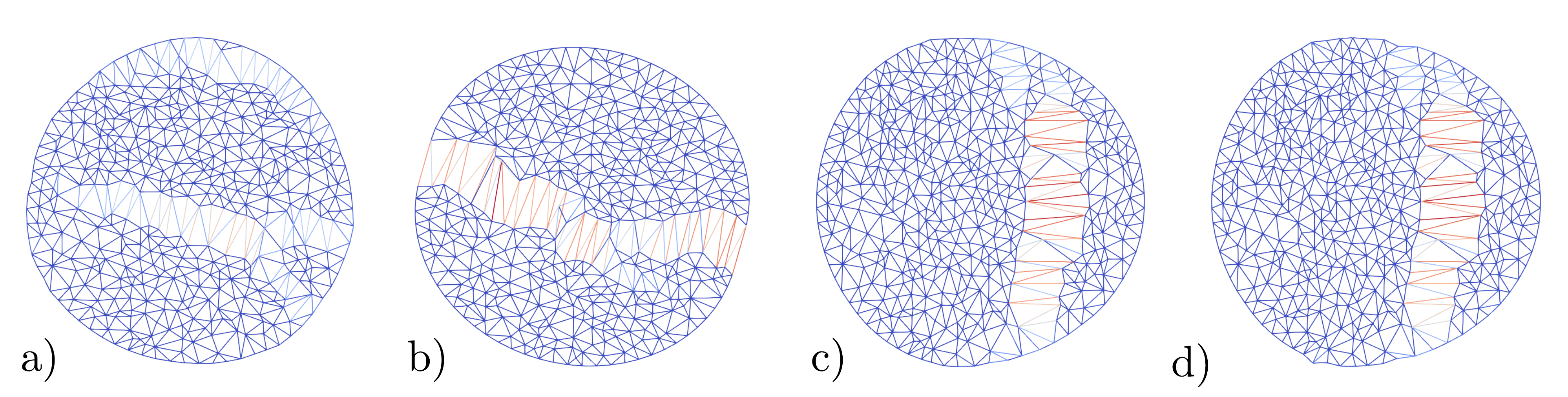}
\caption{Crack patterns in 2D circular fine-scale model of size $d=225$\,mm, loaded by tension under different angles $\alpha$ and boundary conditions: a) $\alpha=1.31$ Periodic; b) $\alpha=1.31$ Embedded-crack; c) $\alpha=0.05$ Periodic; d) $\alpha=0.05$ Embedded-crack. Coloring represents the crack opening, ranging from 0 (blue) to 0.27\,mm (red).\label{fig:2D_circle_cracks}}
\end{figure}

The \emph{circular} fine-scale model, considered only in the smooth surface variant, is generated in 10 different geometries inside a~circle of diameter $d=225$\,mm (its area is equal to the area of the square models). The same analysis is performed, consisting of an~increasing uniaxial tension load in a~set of directions defined by angle $\alpha$. Standard periodic boundary conditions prevent a~localization pattern from crossing the boundary, resulting in the average total energy $W$ at the end of the load being $\sim2-2.5 \times$ higher compared to the reference value obtained in square models, as evidenced in Fig.~\ref{fig:2D_circle_energy_LD}. The same reasoning as before is valid also here: the straining acts over regions of different lengths. Therefore, the energy flux available for crack opening is the largest in the center of the circle and zero at the boundary. The uneven opening of the crack is compensated for by opening the second crack.  Its inability to separate the model into two parts is also manifested in the nonzero residual stress seen in the strain-stress diagram in Fig.~\ref{fig:2D_circle_energy_LD}. The resulting crack patterns for two different load angles are visualized in Fig.~\ref{fig:2D_circle_cracks}a,c, showing that the main crack suffers from uneven opening and another crack might be opened. 

When \emph{Embedded-crack BCs} are applied, the crack can develop evenly across the whole fine-scale model under any loading angle. However, the correct failure mode occurs only if the emerging crack localizes between the opening node pair $a-b$, defined by the vector $\boldv_1$. This situation is shown in Fig.~\ref{fig:2D_circle_cracks}b, while the crack developed in the model of identical internal structure and loading direction with the standard Periodic BCs is shown in Fig.~\ref{fig:2D_circle_cracks}a. When the same model is loaded at different angles $\alpha$, the standard Periodic BCs might lead to a~crack pattern located far from the center (Fig.~\ref{fig:2D_circle_cracks}c). Consequently, also Embedded-crack BCs produce problematic results as the opening node pair is not connected to the main crack (Fig.~\ref{fig:2D_circle_cracks}d). Whether one or the other scenario occurs is a~random process depending on the internal geometry of the model. The consequences of these scenarios on the coarse-scale mechanical response are shown in Fig.~\ref{fig:2D_circle_energy_LD}. The total energy $W$ and stress $S_{\alpha}$ is close to the average reference solution when the localization band occurs between the opening node pair, or close to the average solution with Periodic BCs, when not. Two different internal structures (Geo 1 \& 2) show that the solution randomly oscillates between these scenarios depending on angle $\alpha$. The stress-strain response for one selected angle on the right-hand side shows that correct localization scenario (Geo 2) provides results similar to reference ones, while incorrect scenario (Geo 1) results are similar results with standard Periodic BCs.


\subsection{Fracture in 3D fine-scale models}

All studies of periodic boundary conditions allowing strain localization were done in two dimensions only. Here, the Tessellation and Aligned BC types are implemented in 3D cube models, with the accompanying challenges described in Appendices~\hyperref[appendixA]{A} and \hyperref[appendixB]{B}. Similarly to the 2D study, 10 random fine-scale geometries are generated within a~cube of  size $d=100$\,mm. Only smooth surfaces are considered; models with rough surfaces pose additional challenges for extending the model to 3D, and the Minimal and Aligned BCs remain inapplicable to rough surfaces also in 3D.

\begin{figure}[tb!]
\centering
\includegraphics[width=3.0in]{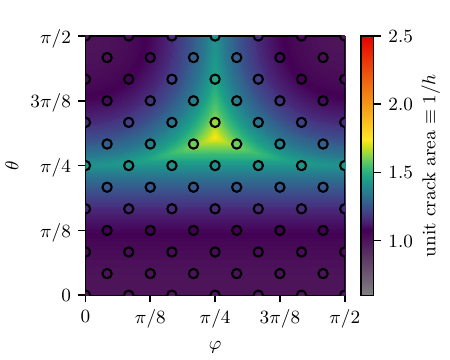}\hfill
\includegraphics[width=3.0in]{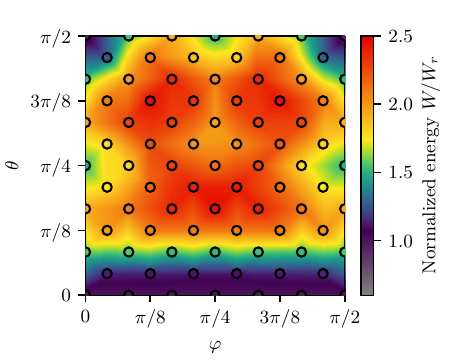}
\caption{Left-hand side: factor $1/h$ expressing theoretical area of inclined crack in a~cube of size 1; right-hand side: average relative energy $W$ computed by cubic fine-scale model of size $d = 100$\,mm with Periodic BCs}
\label{fig:3D_energy_periodic}
\end{figure}

\begin{figure}[tb!]
\centering
\includegraphics[width=3.0in]{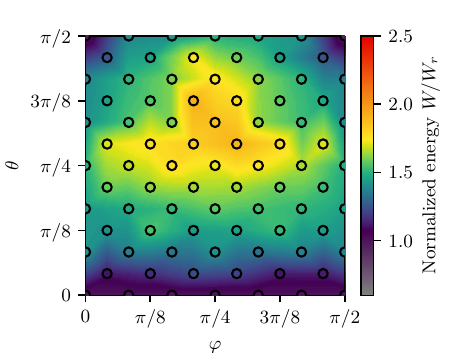}\hfill
\includegraphics[width=3.0in]{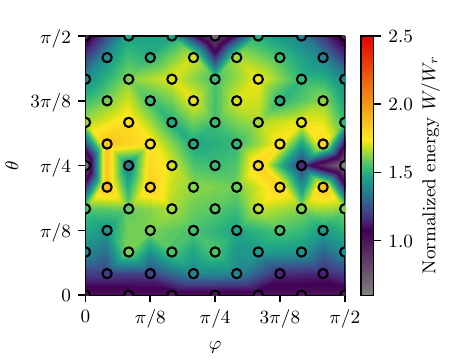}
\caption{Average relative energy $W$ computed by cubic fine-scale model of size $d = 100$\,mm with Periodic BCs: Tessellation BCs (left-hand side) and Aligned BCs (right-hand side)}
\label{fig:3D_energy_tessel}
\end{figure}

\begin{figure}[tb!]
\centering
\includegraphics[width=3.0in]{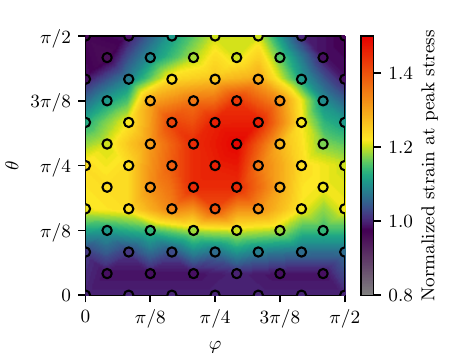}\hfill
\includegraphics[width=3.0in]{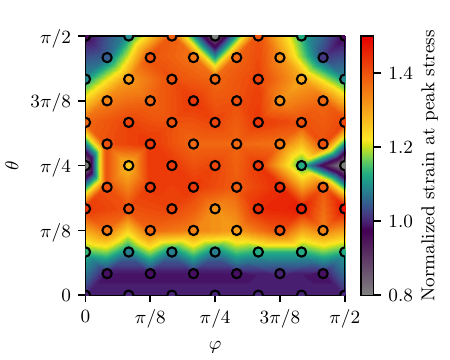}
\caption{Average relative strain at peak stress computed by cubic fine-scale model of size $d = 100$\,mm: left-hand side: Periodic BCs (left-hand side) and Aligned BCs (right-hand side)}
\label{fig:3D_strain}
\end{figure}

Cubic models are loaded by projection of the uniaxial tensile strain tensor in different directions. The normal $\boldn_c$ of the assumed crack surface is considered identical to the direction of the tensile load $\boldv_1$, which is given by the \emph{azimuth angle} $\varphi$ measured from the axis $x_1$ and \emph{the polar} angle $\theta$ measured from the $x_3$ axis. The maximum value of the only nonzero principal strain component $E_I =1.2 \ \times \ 10^{-3}$ is reached in 50 pseudo-time increments. The angles $\varphi$ and $\theta$ vary between $0$ and $\pi/2$, with 85 loading orientations, shown by small circles in Figs.~\ref{fig:3D_energy_periodic}, \ref{fig:3D_energy_tessel} and \ref{fig:3D_strain}.

For the 3D cube, a~scaling factor analogous to that in Eq.~\eqref{eq:h_factor2D} can be derived, resulting in \eqref{eq:h_factor3D}. The inverse of this factor gives the area of a~single planar crack in a~unit cube. The inverses of the three expressions in Eq.~\eqref{eq:h_factor3D} correspond to the area of the projections of the three orthogonal cube faces onto the crack plane defined by the azimuth $\varphi$ and polar $\theta$ angles. The largest of three $h$ values, corresponding to the smallest area, represents the actual area of the crack. The boundaries separating the three cases are obtained by equating the expressions for each pair of cases. 
\begin{align}
h \equiv 
\begin{cases}
\cos \varphi \sin \theta, & \quad 0 \leq \varphi \leq \pi/4 \ \ \  \wedge \ \ \ \theta \leq \arctan(|1/\cos \varphi|) \\
\sin \varphi \sin \theta, & \quad \pi/4 \leq \varphi \leq \pi/2 \ \ \  \wedge \ \ \ \theta \leq \arctan(|1/\sin \varphi|) \\
\cos \theta, & \ \quad  \mathrm{otherwise}
\end{cases}
\label{eq:h_factor3D}
\end{align}
The maximum relative crack area $1/h = 1.732$ is found for $\varphi = \pi/4$ and $\theta = \arccos{1/\sqrt{3}}$. The energy dissipation of a~single crack should be proportional to the area of the crack; therefore, we expect the models to give energy dissipation according to $1/h$ factor. 

\begin{figure}[tb!]
    \centering
    \includegraphics[width=6.5in]{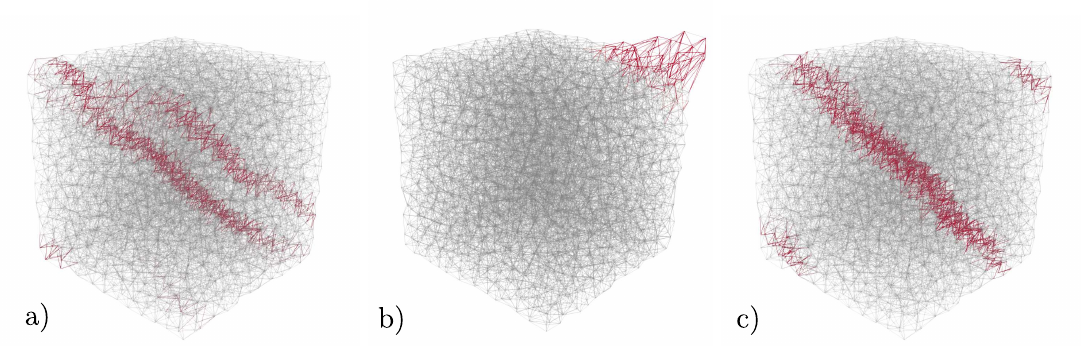}
    \caption{Crack patterns in cubic fine-scale model of size $d=100$\,mm loaded by tension: a) $\varphi=1.05$, $\theta=\pi/4$, Periodic BCs; b) $\varphi=1.31$, $\theta=\pi/4$, Aligned BCs; c) $\varphi=1.05$, $\theta=\pi/4$, Tessellation BCs}
    \label{fig:3D_crack_patterns}
\end{figure}

Fig.~\ref{fig:3D_energy_periodic} shows the total energy $W$ for the \emph{Periodic BCs} normalized by the energy of the \emph{reference} case obtained for $\varphi = 0$ and $\theta = 0$ under the same boundary conditions. The values between the circles, where the results are extracted from the mechanical models, are obtained by linear interpolation. Similarly to the 2D case, energy $W$ provided by the standard Periodic BCs increases for all crack orientations other than those perpendicular to the cube sides. An~example of a~resulting crack pattern with multiple opened cracks driving the increase in energy is shown in Fig.~\ref{fig:3D_crack_patterns}a. The highest energies are about 2.5 larger than in the reference case. The values along the edges $\varphi=0$, $\varphi=\pi/2$, and $\theta=\pi/2$ of the plot correspond to a~loading in which the crack plane is perpendicular to two opposing sides of the cube. These cases should reproduce the behavior of two-dimensional square fine-scale model shown in Fig.~\ref{fig:2D_energy}, and indeed they do. In terms of peak stress, the results are similar to those in 2D, with the peak values being close to the reference solution. However, the strain value corresponding to the peak stress tends to be significantly higher for load orientations close to the diagonal case of  $\varphi = \pi/4$ and $\theta = \arccos{1/\sqrt{3}}$, as seen in Fig.~\ref{fig:3D_strain}.

\begin{figure}[tb!]
    \centering
    \includegraphics[width=6.5in]{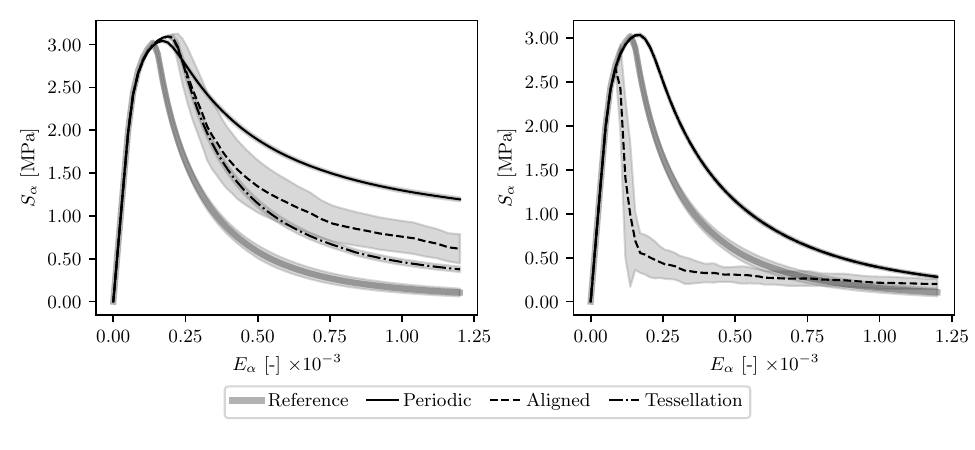}
    \caption{Average stress-strain response and its standard deviation computed from over 10 random geometries of cubic fine-scale model with size $d = 100$\,mm under uniaxial tensile load direction $\varphi=0.26$, $\theta=1.05$ (left-hand side) and $\varphi=\pi/2$, $\theta=\pi/4$ (right-hand side)}
    \label{fig:3D_stress_strain}
\end{figure}

The \emph{Aligned BCs} often lead to energy values closer to those from the reference case, however, for some orientations they exceed it significantly. The reason is the same as in two dimensions; the uneven crack opening due to different strain energy available for cracking results in secondary cracks in the $\Gamma_1$ region. In other cases, the total energy $W$ decreases below the reference value, pointing to spurious localization occurring in the weakly periodic $\tilde{\Gamma}_1$ boundary segment. An~example of such a~crack pattern is shown in Fig.~\ref{fig:3D_crack_patterns}b. Once again, the results of the 2D-like loading seen at the edges $\varphi=0$, $\varphi=\pi/2$, and $\theta=\pi/2$ match the results of 2D simulations (Fig.~\ref{fig:2D_energy}), apart from the central values, equivalent to $\alpha=\pi/4$ case in 2D, where in 3D spurious localization lowers $W$ below the reference value. The peak stress is close to the reference value for all loading orientations, except for the cases of spurious localization discussed earlier. This contrasts with the 2D case, where most orientations yielded substantially lower values (Fig.~\ref{fig:2D_stress_peak}). The likely cause is the stabilization of the model by the third direction of constraints. Probably for the same reason, the strain values corresponding to the peak stress state are also below the reference value only in spurious localization cases and otherwise are higher, as seen in Fig.~\ref{fig:3D_strain}.
The \emph{Aligned BCs} also lead to a~higher variance in response across the different geometries in the model set compared to other BC variants. This can be seen in the stress-strain diagrams in Fig.~\ref{fig:3D_stress_strain}, where the standard deviation is plotted around the mean response of 10 internal structures. Diagrams for two loading directions are shown: the one on the left-hand side resulted for Aligned BCs in a~single localization band, while the other one shows a~case of spurious localization.

The \emph{Tessellation boundary conditions} show promising results also in the three-dimensional case. Once again, they produce a~single crack for all load orientations. An~example of a~single crack, leaving and re-entering the cube model, resulting in three crack segments is shown in Fig.~\ref{fig:3D_crack_patterns}c. As seen in Fig.~\ref{fig:3D_energy_tessel}, the energy $W$ depends on the orientation of the crack in a~way consistent with the $1/h$ factor, with a~maximum value close to the orientation of the diagonal crack at $\varphi = \pi/4$ and $\theta = \arccos{1/\sqrt{3}}$. The results do not match the factor $1/h$ as well as in the 2D case, which is likely caused by the discretization of the boundary conditions and the small size of the model. In terms of peak stress and the corresponding strain values, the results are similar to those of the \emph{Aligned BCs} shown earlier, but there is no spurious localization. The variance of the response across different internal structures is similar to the Periodic BCs, see Fig.~\ref{fig:3D_stress_strain}.

The results of the \emph{Minimal kinematic BCs} and the weaker version of \emph{Aligned BCs} are not presented. They are expected to also suffer from spurious localization in three dimensions. Similarly, the defects of the \emph{Embedded-crack BCs} constraining the circular model are expected to be manifested also in spherical models.

\subsection{Constraint complexity and computational cost}
The computational cost associated with the investigated boundary conditions can be divided into two components: constraint generation and their influence on the iterative equilibrium calculation.

The standard periodic constraints are directly defined by the model geometry and require no additional preprocessing. The generation of the Aligned and Tessellation BCs can be costly, especially in 3D, depending on the number of boundary nodes and the projection tessellation cell size (Appendix~\hyperref[appendixA]{A}). The Tessellation BCs are advantageous in this regard because one of the boundary projections retains the predefined periodic constraints. In our case with predefined crack normal, the constraints are generated only once and prior to the simulation run. Moreover, the associated computational cost is negligible compared to the cost of the simulation itself.

\begin{figure}[tb!]
\centering
\includegraphics[width=3.2in]{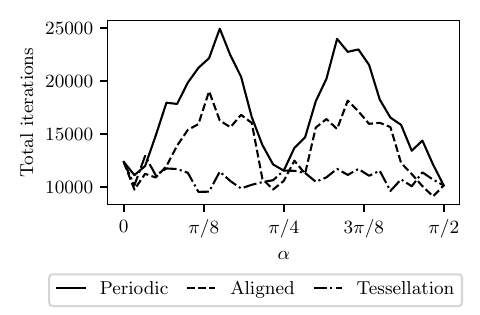}\hfill
\includegraphics[width=3.2in]{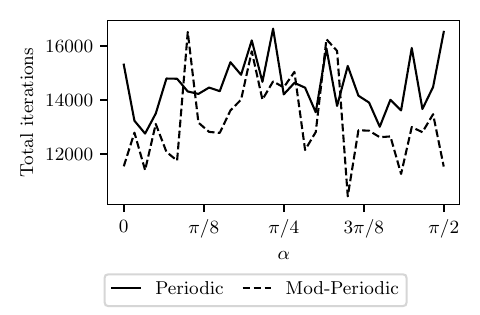}
\caption{Average total iterations during uniaxial loading simulation; left-hand side: smooth-surface square models; right-hand side: circular models}
\label{fig:2D_iterations}
\end{figure}

In the square and cubical models, the constraints are imposed in the form of Lagrange multipliers. The standard periodic BCs represent the highest number of constraints, while the other variants result in the same or lower number, equal to the number of the integration zones. The maximum differences in the number of constraints are roughly around 20\,\%, therefore, the computational cost per iteration is comparable among all variants. The total computational cost therefore primarily changes with the number of iterations required by the Newton-Raphson solver. Fig.~\ref{fig:2D_iterations} shows that only the Tessellation BCs lead to identical number of iterations irrespectively of the loading direction thanks to the unobstructed development of single straight localization band.

For the circular model, introducing a displacement jump into the periodic boundary conditions does not increase the number of constraints and any change in per-iteration cost is negligible. Fig.~\ref{fig:2D_iterations} shows that no significant difference between the BC variants was found in terms of average number of total iteration over the model sets. 

\section{Conclusions}
Three recently proposed periodic boundary condition (BC) formulations, designed to mitigate the load-orientation dependence of the homogenized response of softening fine-scale models, were implemented for discrete mechanical models. The formulations of the Percolation-path-aligned BC and Tessellation BC variants were extended to three-dimensions, with the implementation procedures presented in detail.

An analysis of the macroscopic response and localization patterns of square, cubic, and circular fine-scale discrete particle models loaded in uniaxial tension with various forms of boundary conditions was performed, including novel BC variants and traditional BC formulations. The following conclusions are drawn.

\begin{itemize}
\item Standard \emph{Periodic} boundary conditions (BCs) produce a~strongly orientation dependent softening response of the fine-scale model. The Periodic BCs lead to correctly developed localization band only for a~few load orientations in the square and cube models, in circular models they forbid proper band development for all load orientations. The \emph{Minimal kinematic} \parencite{minimalBC} cannot be used with softening models at all due to the spurious localization into few boundary elements. 

\item The \emph{Percolation-path-aligned BCs} \parencite{CoeKouGee12}, which rotate the periodicity frame with the localization band, have been applied to the square and cube models. It was shown that for a~certain range of load orientations, the difference in straining between its weakly and strongly periodic segments leads to multiple localization bands, which cause overestimation of dissipated energy and overly ductile behavior. This was evidenced by the introduction of the \emph{weak} modification, which replaces the strongly periodic segment with a the weakly periodic one. This variant does not suffer from the aforementioned issue, however, because the weakly constrained segment can be prone to spurious localization observed with Minimal kinematic BCs, this variant is also not viable. While both these issues might not manifest for different geometries or material parameters, they are inherent to these boundary conditions.

\item The \emph{Tessellation BCs} \parencite{goldmann2018boundary}, which shift the periodicity frame so that it can accommodate inclined localization bands without periodic repetitions, have proven to offer a~consistent response when applied to square models with uniform and periodic boundaries, as well as to cubic models. A~single localization band, which may exit and reenter the model, develops for any load orientation. The dissipated energy in two dimensions is proportional to the crack area, which depends solely on the geometry of the domain and can be easily calculated, allowing for potential post-processing corrections. In three dimensions, the proportionality is weaker but still reasonable. The Tessellation BCs were found to offer a~significant improvement over standard Periodic BCs.

\item The \emph{Embedded-crack periodic BCs} \parencite{sperical1}, which introduce a~displacement jump in the boundary conditions for circular domains, allow a~full localization band separating two halves of the model. However, this is conditioned by the initiation of the crack in the central part of the circle. The crack may then connect with the displacement jump at the boundary. When the crack initiates outside the domain center, these boundary conditions produce multiple localization bands and dissipate an~excessive amount of energy.
\end{itemize}

The present conclusions apply to the discrete models considered here, which assume homogeneous material parameters and monotonic uniaxial tensile loading. In other types of models, such as continuum models with a~heterogeneous distribution of material parameters, additional complications might appear (for example, stiff inclusions combined with shifted periodicity might prevent strain localization~\parencite{shifting_of_microstructure}). Furthermore, the applicability of the studied boundary conditions in more complex loading scenarios is uncertain. Further work would require a robust technique to determine the crack normal at the onset of localization, which sets the rotated periodicity frame and therefore constraints further crack propagation. 

\appendix
\section*{Appendix A: Discretization of Boundary Projections and Constraints Definition} \label{appendixA}

\begin{figure}[tb!]
    \centering
    \includegraphics[width=5.6in]{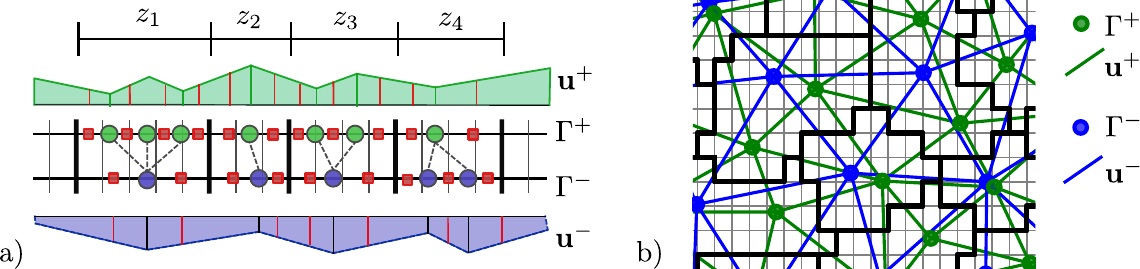}
    \caption{a) Boundary nodes on + (green) and - (blue) side of 1D projection in 2D models, connected into groups (dashed lines), around which the integration zones $z_a$ (thick black lines) are built from cells (thin gray lines). The linear displacement field $\boldu^+$ and $\boldu^-$ are numerically integrated in the integration points (red); b) the same principle applied in a~2D projection of a~3D 
    model, integration points are hidden to improve image clarity}
    \label{fig:2D3D_zones}
\end{figure}
The periodicity of the fine-scale model boundary nodes does not conform to the rotated periodic frame that must be imposed in many of the boundary conditions studied. To accommodate the rotated periodic frame, both $\Gamma^+$ and $\Gamma^-$ are partitioned into zones. Each such zone contains \textit{exactly} one node on one side (either $+$ or $-$) and \textit{at least} one node on the opposite side. This ensures that all nodes are constrained and that the number of constraints is as close as possible to the number of constraints in the periodic model. 

Algorithmically, the zones are created by the following procedure. Firstly, \emph{groups} of $+$ and $-$ nodes are found. Each node on the $+$ side finds the closest (in the projection) node on the $-$ side, denoted the central node (possibly more $+$ nodes are attached to the same central node). Then, each $-$ node that was not previously connected to a $+$ node finds the closest $+$ node. This can lead to clusters with multiple $+$ and $-$ nodes. To separate such a~group, the links between the central node and the $+$ nodes are kept only if the $+$ node is not connected to any other $-$ node. In case no such $+$ node exists, only the shortest link is kept.

With the groups completed, the projection area is divided into a~number of \textit{cells}. In 2D models, these cells are small line sections of equal size, in 3D models small rectangular cells are used. Then each cell is assigned to the closest node in the projection (not distinguishing the $+$ and $-$ sides). Cells that belong to a~single connected group of nodes then constitute the integration zone $a$. 

The strong periodicity of translations and rotations in zone $a$ is prescribed by the integral condition
\begin{align}
\int_{\Gamma_a^+}\boldu \mathrm{d}\Gamma_a^+ &= \int_{\Gamma_a^-}\boldu \mathrm{d}\Gamma_a^-  & \int_{\Gamma_a^+}\boldtheta \mathrm{d}\Gamma_a^+ &= \int_{\Gamma_a^-}\boldtheta \mathrm{d}\Gamma_a^- 
\end{align}
The integral is evaluated numerically. In 2D, linear interpolation of displacements between nodes is assumed, and integration points lie in the center between two nodes or nodes and the zone boundary. This is depicted in Fig.~\ref{fig:2D3D_zones}a. A~simpler variant is used in 3D models. The integration points lie in the centers of the cells and their displacement value is taken from the closest node. This approximate constraint was verified for elastic material behavior, where it yields practically identical results as the standard periodic models aligned with the periodic frame.

Note that if only one zone is considered for the whole boundary, the integral periodic boundary conditions become the Minimal kinematic BCs, except of course the rotations that are left unconstrained under Minimal kinematic BCs.  

The constraints are prescribed in the form of Lagrange multipliers, and a~direct solver is used to ensure that they are exactly satisfied.

\section*{Appendix B: Choice of Directions Parallel to the Crack in 3D}\label{appendixB}
\begin{figure}[tb!]
    \centering
    \includegraphics[width=6.3in]{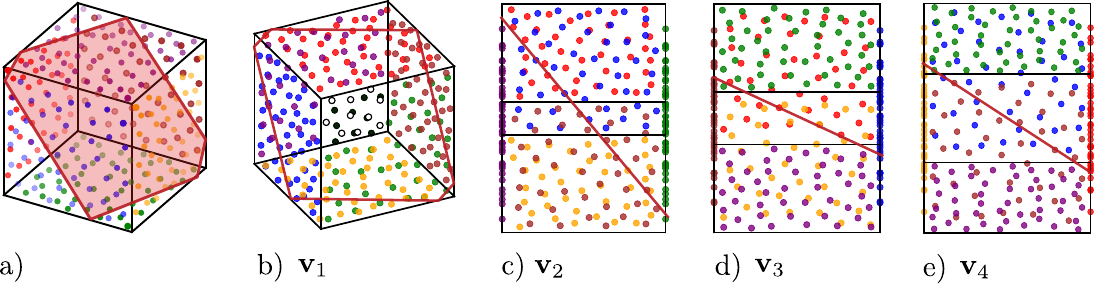}
    \caption{Boundary nodes a~cubical fine-scale model; a) 3D view with a~possible crack location and shape in Aligned BCs; b) through-crack projection along the $\boldv_1$ vector,  with black and white nodes lying in the $\Gamma_1$ boundary section and the rest in the $\tilde{\Gamma}_1$ section; c), d), e) parallel to crack projections along $\boldv_2$, $\boldv_3$, $\boldv_4$ vectors}
    \label{fig:3D_projections}
\end{figure}

Three projection directions must be selected for the Aligned BCs when applied to the cubic domain, so that it corresponds to the Periodic BCs when crack normal is perpendicular to the cube sides. The first ($\boldv_1$) is given by the loading direction, the other two ($\boldv_2$ and $\boldv_3$) might be constrained to create orthonormal system with the first one ($\boldv_1 \perp \boldv_2 \perp \boldv_3$) , but there are still infinitely many options for them. 

In this work, $\boldv_2$ and $\boldv_3$ are selected such that the two opposing sides are projected to a~line along each of them. Therefore, these sides are omitted when defining the constraints in the $\boldv_2$ and $\boldv_3$ directions. The total number of constraints decreases and becomes similar to that of periodic BCs. Such directions can be obtained as a~cross product of the crack normal $\boldn_c=\boldv_1$ and the model's side normals, here coincident with the unit vectors of a~Cartesian basis $\bolde_1$, $\bolde_2$, and $\bolde_3$. This yields three directions $\boldv_2,\boldv_3,\boldv_4$. These projections are shown in Fig.~\ref{fig:3D_projections}. Two of these three parallel-to-crack directions need to be chosen so that the total number of projections is three. The most convenient pair was chosen to minimize the difference in the density of nodes on the $+$ and $-$ parts of the boundary. The uneven densities of the nodes degrade the quality of the periodic constraint imposed on the opposite $\pm$ boundary parts. Technically, this is done by dropping the vector that is closest to some of the cube side normals. Note that this means that $\boldv_2$ and $\boldv_3$ are generally only orthogonal with $\boldv_1$ and not with each other, however, this was found to not be an issue. Additionally, in case of crack normal perpendicular to the cube sides, $\boldv_1$, $\boldv_2$ and $\boldv_3$ correspond to $\bolde_1$, $\bolde_2$, and $\bolde_3$ (not necessarily in this order), leading to Aligned BCs corresponding with Periodic BCs.

\section*{Acknowledgement}
The authors gratefully acknowledge financial support from the Czech Science Foundation under project number GA24-11845S. The simulations were performed  on computational servers acquired within project INODIN (Innovative methods of materials diagnostics and monitoring of engineering infrastructure to increase its durability and service time – CZ.02.01.01/00/23\_020/0008487) co-funded by European Union. The authors also acknowledge inspiring discussions with Milan Jirásek and Martin Doškář from Czech Technical University.

\section*{Data availability}
The input files for OAS solver and results presented in this article, together with Python scripts used for generation of the Aligned and Tessellation BCs, are available at \url{https://doi.org/10.5281/zenodo.20415173}.

\printbibliography

\break
\end{document}